%% file: ms.tex
\documentclass[10pt,journal,compsoc]{IEEEtran}

\input{macro}

\usepackage{multirow}
\usepackage{lscape}
\usepackage{graphicx}
\usepackage{url}
\usepackage{framed}
\usepackage[usenames, dvipsnames]{color}
\usepackage{balance}
\usepackage{colortbl}
\usepackage{nopageno}
\newcommand{\blue}[1]{\textcolor{black}{#1}}
\newcommand{\apfd}{APFD}
\newcommand{\apfdc}{APFDc}

\makeatletter
  \newcommand{\miniscule}{\@setfontsize\miniscule{3}{4}}
  \newcommand{\minisculee}{\@setfontsize\minisculee{4}{5}}
\makeatother

\begin{document}

\title{How Do Static and Dynamic Test Case Prioritization Techniques Perform
on Modern Software Systems?  An Extensive Study on GitHub Projects}

\author{Qi~Luo,~\IEEEmembership{Student Member,~IEEE,}
        Kevin~Moran,~\IEEEmembership{Student Member,~IEEE,}
        Lingming~Zhang,~\IEEEmembership{Member,~IEEE,}
        and~Denys~Poshyvanyk,~\IEEEmembership{Member,~IEEE}
\IEEEcompsocitemizethanks{
\IEEEcompsocthanksitem Qi Luo, Kevin Moran, and Denys Poshyvanyk are with the Department
of Computer Science, College of William and Mary, Williamsburg, VA, 23188.\protect\\
E-mail: \{qluo,kpmoran,denys\}@cs.wm.edu
\IEEEcompsocthanksitem Lingming Zhang is with the Department of Computer Science, University of Texas at Dallas, Dallas, TX, 75080. \protect\\
E-mail: lingming.zhang@utdallas,edu}
\thanks{This paper is an extension of ``A Large-scale Empirical Comparison of Static and Dynamic Test Case Prioritization Techniques'' that appeared in Proceedings of 24th ACM SIGSOFT International Symposium on the Foundations of Software Engineering (FSE'16), Seattle, WA, November 13-18, 2016, pp. 559-570. \protect\\
Manuscript received May, 2017.}}

\markboth{IEEE Transactions on Software Engineering,~Vol.~x, No.~x, ~2017}%
{}

\IEEEtitleabstractindextext{%
\begin{abstract}

Test Case Prioritization (TCP) is an increasingly important regression testing technique for reordering test cases according to a pre-defined goal, particularly as agile practices gain adoption. To better understand these techniques, we perform the first extensive study aimed at empirically evaluating four static TCP techniques, comparing them with state-of-research dynamic TCP techniques across several quality metrics. This study was performed on 58 real-word Java programs encompassing 714 KLoC and results in several notable observations.  First, our results across two effectiveness metrics (the Average Percentage of Faults Detected \textit{APFD} and the cost cognizant \textit{APFDc}) illustrate that at test-class granularity, these metrics tend to correlate, but this correlation does not hold at test-method granularity.  Second, our analysis shows that static techniques can be surprisingly effective, particularly when measured by APFDc. \blue{Third, we found that TCP techniques tend to perform better on larger programs, but that program size does not affect comparative performance measures between techniques. Fourth, software evolution does not significantly impact comparative performance results between TCP techniques. Fifth,} neither the number nor type of mutants utilized dramatically impact measures of TCP effectiveness under typical experimental settings.  Finally, our similarity analysis illustrates that highly prioritized test cases tend to uncover dissimilar faults.

\end{abstract}

}

\maketitle

\IEEEdisplaynontitleabstractindextext

\IEEEpeerreviewmaketitle

\input{intro}

\input{background}

\input{study}

\input{result}

\input{threats}

\input{lessons}

\input{conclude}

\input{acknowledgments}



\bibliographystyle{abbrv}
\bibliography{ms}
\vspace{-2cm}
\begin{IEEEbiography}[{\includegraphics[width=1in,height=1.25in,clip,keepaspectratio]{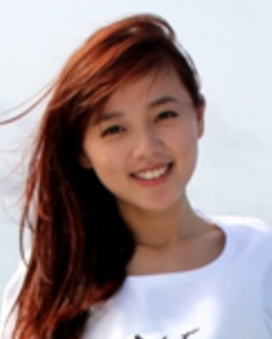}}]{Qi Luo}
is a Ph.D. candidate in the department of Computer Science at the College of William and Mary. She is a member of SEMERU research group and advised by Dr. Denys Poshyvanyk. Her research interests are in software engineering, performance testing and regression testing. She received her Bachelor degree from Beihang University and her Master degree from Tsinghua University. She has published in several top software engineering conferences including: FSE, ICSE and MSR. She is a student member of ACM and IEEE. 
\end{IEEEbiography}
\vspace{-1.5cm}
\begin{IEEEbiography}[{\includegraphics[width=1in,height=1.25in,clip,keepaspectratio]{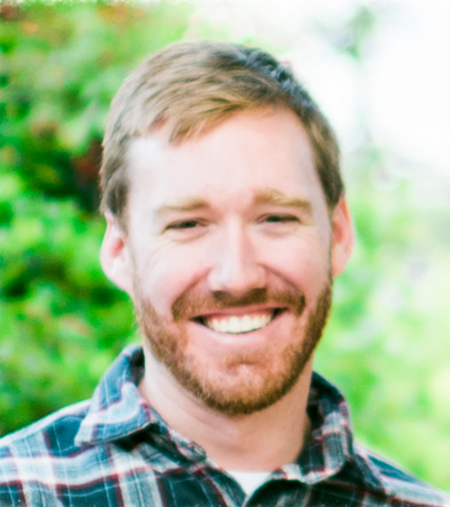}}]{Kevin Moran}
is currently a Ph.D student in the Computer Science Department at the College of William and Mary. He is a member of the SEMERU research group and advised by Dr. Denys Poshyvanyk. His main research interest involves facilitating the processes of Software Engineering, Maintenance, and Evolution with a focus on mobile platforms. He graduated with a M.S. degree from William and Mary in August of 2015 and his thesis focused on improving bug reporting for mobile apps through novel applications of program analysis techniques. He has published in several top peer-reviewed software engineering venues including: ICSE, ESEC/FSE, ICSME, ICST, and MSR. He was recently recognized as the second-overall winner among graduate students in the ACM Student Research competition at ESEC/FSE15. Moran is a student member of IEEE and ACM and has served as an external reviewer for ICSE, ICSME, FSE, APSEC, and SCAM. More information available at https://www.kpmoran.com.
\end{IEEEbiography}
\vspace{-1.5cm}
\begin{IEEEbiography}[{\includegraphics[width=1in,height=1.25in,clip,keepaspectratio]{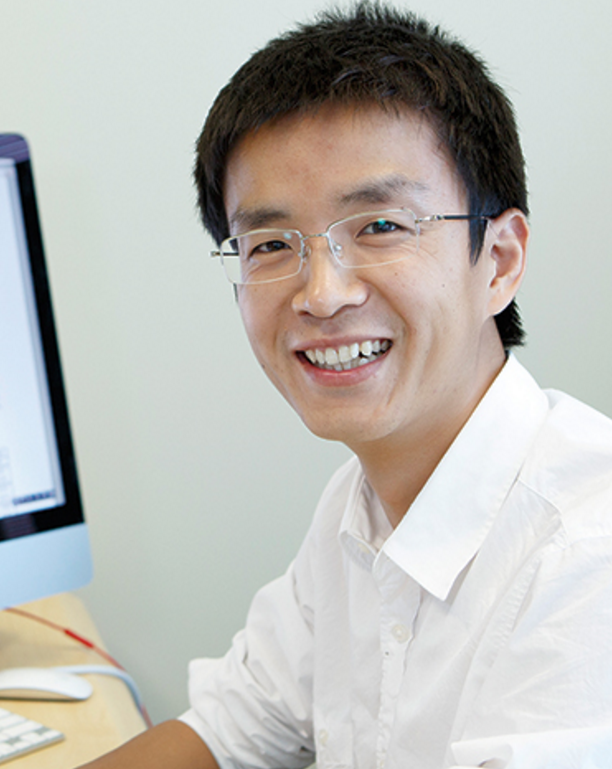}}]{Lingming Zhang} is an assistant professor in the Computer Science Department at the University of Texas at Dallas. He obtained his Ph.D. degree from the Department of Electrical and Computer Engineering in the University of Texas at Austin in May 2014.  He received his MS degree and BS degree in Computer Science from Peking University (2010) and Nanjing University (2007), respectively. His research interests lie broadly in software engineering and programming languages, including automated program analysis, testing, debugging, and verification, as well as software evolution and mobile computing. He has authored over 30 papers in premier software engineering or programming language transactions and conferences, including ICSE, FSE, ISSTA, ASE, POPL, OOPSLA, TSE and TOSEM. He has also served on the program committee or artifact evaluation committee for various international conferences (including ASE, ICST, ICSM, ISSRE, OOPSLA, and ISSTA). His research is being supported by NSF and Google. More information available at: http://www.utdallas.edu/~lxz144130/.
\end{IEEEbiography}
\vspace{-2cm}
\begin{IEEEbiography}[{\includegraphics[width=1in,height=1.25in,clip,keepaspectratio]{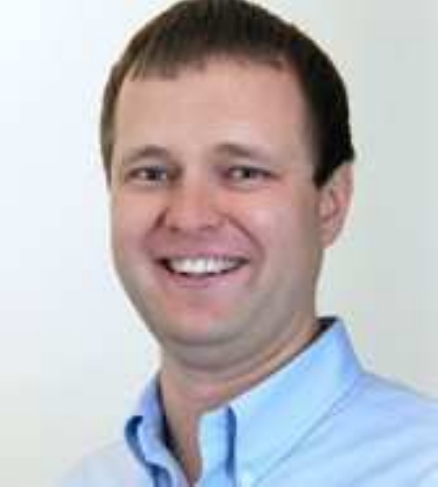}}]{Denys Poshyvanyk} is the Class of 1953 Term Associate Professor of Computer Science at the College of William and Mary in Virginia. He received the MS and MA degrees in Computer Science from the National University of Kyiv-Mohyla Academy, Ukraine, and Wayne State University in 2003 and 2006, respectively. He received the PhD degree in Computer Science from Wayne State University in 2008. He served as a program co-chair for ICSME'16, ICPC'13, WCRE'12 and WCRE'11. He currently serves on the editorial board of IEEE Transactions on Software Engineering (TSE), Empirical Software Engineering Journal (EMSE, Springer) and Journal of Software: Evolution and Process (JSEP, Wiley). His research interests include software engineering, software maintenance and evolution, program comprehension, reverse engineering, software repository mining, source code analysis and metrics. His research papers received several Best Paper Awards at ICPC'06, ICPC'07, ICSM'10, SCAM'10, ICSM'13 and ACM SIGSOFT Distinguished Paper Awards at ASE'13, ICSE'15, ESEC/FSE'15, ICPC'16 and ASE'17. He also received the Most Influential Paper Awards at ICSME'16 and ICPC'17. He is a recipient of the NSF CAREER award (2013).  He is a member of the IEEE and ACM. More information available at: http://www.cs.wm.edu/~denys/.
\end{IEEEbiography}

\end{document}

%% file: intro.tex
\IEEEraisesectionheading{\section{Introduction}}
\label{sec:intro}

\IEEEPARstart{M}{\blue{odern}} \blue{ software evolves at a constant and rapid pace; developers continually add new features and fix bugs to ensure a satisfied user base. During this evolutionary process, it is crucial that developers do not introduce new bugs, known as software \textit{regressions}.} {\em Regression testing} is a methodology for efficiently and effectively validating software changes against an existing test suite aimed at detecting such bugs \cite{Lu:ICSE16,Zhang:ICSM11}. \blue{One of the key tasks of the contemporary practice of continuous regression testing, is test case prioritization (TCP).} 

Regression test prioritization techniques reorder test executions in order to maximize a certain objective function, such as exposing faults earlier or reducing the execution time cost \cite{Lu:ICSE16}. \blue{This practice can be readily observed in applications to large industrial codebases such as at Microsoft, where researchers have built test prioritization systems for development and maintenance of Windows for a decade~\cite{Srivastava:ISSTA02, Czerwonka:ICST11}. In academia, there exists a large body of research that studies the design and effectiveness of regression TCP techniques~\cite{Walcott:06,zhang2013bridging,Rothermel:01,Rothermel:99, Lu:ICSE16, Korel:05}. Traditionally, TCP techniques leverage one of several code coverage measurements of tests from a pervious software version as a representation of test effectiveness on a more recent version. These approaches use this coverage-based test adequacy criterion to iteratively compute each test's priority, and then rank them to generate a prioritized list. Researchers have proposed various forms of this traditional approach to TCP, including greedy (total and additional strategies)
~\cite{zhang2013bridging,Rothermel:01,Rothermel:99}, adaptive random testing ~\cite{Jiang:09}, and search-based strategies~\cite{Li:07}.}

\blue{While dynamic TCP techniques can be useful in practice, they may not be always applicable due to certain notable shortcomings, including: 1) the time cost of executing an instrumented program to collect coverage information~\cite{gligoric2015practical,Mei:TSE12}; 2) expensive storage and maintenance of coverage information~\cite{Mei:TSE12,Zhang:ICSM09}; 3) imprecise coverage metrics due to code changes during evolution or thread scheduling of concurrent systems~\cite{legunsen2016extensive}, and 4) the absence of coverage information for newly added tests~\cite{Lu:ICSE16} or systems/modules that disallow code instrumentation~\cite{legunsen2016extensive} (e.g., code instrumentation may break the time constraints of real-time systems). Thus, to offer alternative solutions that that do not exhibit many of these shortcomings, researchers have proposed a number of TCP techniques that rely solely upon \textit{static} information extracted from the text of source and test code.  Unfortunately, since the introduction of purely static TCP techniques, little research has been conducted to fully investigate the effectiveness of static techniques on modern software. This begs several important questions in the context of past work on dynamic techniques, such as:} \textit{How does the effectiveness of static and dynamic techniques compare on modern software projects?} \textit{Do static and dynamic techniques uncover similar faults?} \textit{How efficient are static techniques when compared to one another?}  \blue{The answers to these questions are of paramount importance as they will guide future research directions related to TCP techniques.}

	Several empirical studies have been conducted in an attempt to examine and understand varying aspects of different TCP approaches~\cite{Rothermel:99,Elbaum:TSE02,Do:04,Qu:ISSTA08,Thomas:EMSE14}.  \blue{However, there are clear limitations of prior studies that warrant further experimental work on TCP techniques: 1) recently proposed TCP techniques, particularly static techniques, have not been thoroughly evaluated against each other or against techniques that operate upon dynamic coverage information; 2) no previous study examining static TCP approaches has comprehensively examined the impact of different test granularities (e.g., prioritizing entire test classes or individual test methods), the efficiency of the techniques, or the similarities in terms of uncovered faults; 3) prior studies have typically failed to investigate the application of TCP techniques to sizable real-world software projects, \blue{and none of them have investigated the potential impact of program size (i.e., LOC) on the effectiveness of TCP techniques};} 4) prior studies have not comprehensively investigated the impact of the quantities of faults used to evaluate TCP approaches; and \blue{5) no previous study has attempted to gain an understanding of the impact of fault characteristics on TCP evaluations.} 

	Each of these points are important considerations that call for thorough empirical investigation. \blue{For instance, studying the effectiveness and similarity of faults uncovered for \textit{both} static and dynamic techniques could help inform researchers of potential opportunities to design more effective and robust TCP approaches. Additionally, evaluating a set of popular TCP techniques on a large group of sizable real-world java programs would help bolster the generalizability of performance results for these techniques. Another important consideration that arises from limitations of past studies is that an increasing number of studies use mutants as a proxy for real faults to evaluate performance characteristics of TCP techniques. Thus, understanding the effect that mutant quantities and operators have on mutation analysis-based TCP evaluations should help researchers design more effective and reliable experiments, or validate existing experimental settings for continued use in future work.  Therefore, in this paper we evaluate the effectiveness of TCP approaches in terms of detecting mutants.}

\blue{To answer the unresolved questions related to the understanding of TCP techniques and address the current gap in the existing body of TCP research we perform an extensive empirical study comparing four popular static TCP techniques}, i.e., call-graph-based (with total and additional strategies) \cite{Zhang:ICSM09}, string-distance-based \cite{Ledru:ASE12}, and topic-model based techniques \cite{Thomas:EMSE14} to four state-of-the-art dynamic TCP techniques (i.e., the greedy-total \cite{Rothermel:99}, greedy-additional \cite{Rothermel:99}, adaptive random \cite{Jiang:09}, and search-based techniques \cite{Li:07}) on 58 real-world software systems. \blue{All of the studied TCP techniques were implemented based on the papers that initially proposed them and the implementation details are explained in Section \ref{sec:impl}.  It is important to note that different granularities of dynamic coverage information may impact the effectiveness of dynamic TCP techniques. In this paper, we examine statement-level coverage for dynamic techniques, since previous work \cite{Lu:ICSE16,Mei:TSE12} has illustrated that statement-level coverage is \textit{at least as effective} as other common coverage criteria (e.g., method and branch coverage) in the TCP domain. In our evaluation criteria we examine the effectiveness of the studied techniques in terms of the Average Percentage of Faults Detected (APFD) and its cost cognizant version APFDc. Additionally, we analyze the implications of these two metrics as efficacy measures of TCP techniques and discuss the implications of this analysis. We also analyze the impact of subject size and software evolution on these metrics. Furthermore, during our empirical study, we vary the operator types and the quantities of injected mutants to investigate whether these factors significantly affect the evaluation of TCP approaches. We also examine the similarity of detected of faults for the resultant prioritized sets of test cases generated by our studied TCP techniques at different test granularities (e.g., both method and class levels). More specifically, we investigate the total number and the relative percentages of different types of mutants detected by the most highly prioritized test cases for each TCP technique to further understand their capabilities in detecting faults with varying attributes. Finally, we examine the efficiency, \blue{in terms of execution time (i.e., the processing time for TCP technique), of static TCPs to better understand the time cost associated with running these approaches.}}

\blue{Our study bears several notable findings.  The first of these is that \textit{there are statistically significant differences among the APFD values of all studied techniques.} When measuring the average APFD values across our subject programs, we found that the call-graph-based (with ``additional" strategy) technique outperforms all studied techniques at the test-class level.} At the test-method level, the call-graph and topic-model based techniques perform better than other static techniques, but worse than two dynamic techniques, i.e., the additional and search-based techniques.   

	Second, \textit{our results demonstrate that \blue{APFDc} values are generally consistent with APFD values at test-class level but relatively less consistent at test-method level.} When examining the effectiveness of TCP approaches in terms of the cost-cognizant APFDc values, we found that the call-graph-based (with ``additional" strategy) technique outperforms all studied dynamic and static techniques at both test-class and test-method levels, indicating the limitations of dynamic execution information in reducing actual regression testing time costs.  \blue{Additionally, while APFDc values vary dramatically across 58 subject programs, based on the results of our analysis, there are no statistically significant differences between TCP techniques based on APFDc values at both of test-class and test-method level when controlling for the subject program.} 

	Third, \textit{our experiments indicate that the test granularity dramatically impacts the effectiveness of TCP techniques}. While nearly all techniques perform better at method-level granularity based on both of APFD and APFDc values, the static techniques perform comparatively worse to dynamic techniques at method level as opposed to class level based on APFD values.

	Fourth, \textit{our study shows that differences in the subject size, software version, mutant types, or mutant quantities tend not to largely impact experimental results measuring TCP performance.} In terms of execution time, call-graph based techniques are the most efficient of the static TCP techniques. 

	Finally, \textit{the results of our similarity analysis study suggest that there is minimal overlap between the uncovered faults of the studied dynamic and static TCPs, with the top 10\% of prioritized test-cases only sharing $\approx$ 25\% - 30\% of uncovered faults.} Thus, the most highly prioritized test cases from different TCP techniques exhibit dissimilar capabilities in detecting different types of mutants. This suggests that certain TCP techniques may be better at uncovering faults (or mutants) that exhibit certain characteristics, and that aspects of different TCP techniques may be combined together to alter performance characteristics.  Both of these findings are promising avenues for future work.

\blue{To summarize, this paper makes the following noteworthy contributions summarized in Table~\ref{tab:cont}.
\begin{table}[h!]
\blue{
\caption{\label{tab:cont}The List of Contributions}
\begin{tabular}{|p{3.5cm}|p{4.5cm}|}
\hline
Contributions & Descriptions\\\hline\hline
{\bf Static vs. Dynamic TCP} & To the best of the authors' knowledge, this is the first extensive empirical study that compares the effectiveness, efficiency, and similarity of uncovered faults of both static \textit{and} dynamic TCP techniques on a large set of modern real-world programs;\\ \hline
{\bf Impact of Performance Metrics} &We evaluate the performance of TCP techniques based on two popular metrics, APFD and APFDc, and understand the relationship between the performance of these two metrics for TCP evaluation;\\ \hline
{\bf Impact of Test Case Granularity}&We evaluate the performance of TCP techniques at two different test granularities, and investigate the impacts of test granularities on TCP evaluation;\\ \hline
{\bf Impact of Program Subject Size}&{We evaluate the impacts of subject size on the effectiveness of the studied static and dynamic TCP techniques;} \\ \hline
{\bf Impact of Software Evolution}&{We evaluate the impacts of software evolution on the effectiveness of the studied static and dynamic TCP techniques;} \\ \hline
{\bf Impact of the Number of Studied Faults}&{We conduct the first study investigating the impact of different fault quantities used in the evaluation on the effectiveness of TCP techniques;}\\ \hline
{\bf Impact of Fault Types}&{We conduct the first study investigating the impact of different fault types used in the evaluation on the effectiveness of TCP techniques;}\\ \hline
{\bf Analysis of Similarity of Uncovered Faults}&{We analyze the similarity of faults uncovered between prioritized sets of test cases generated by different techniques;}\\ \hline
{\bf Practical Guidelines for Future Research}&{We discuss the relevance and potential impact of the findings in the study, and provide a set of learned lessons to help guide future research in TCP;}\\ \hline
{\bf Open Source Dataset}&We provide a publicly available, extensive online appendix and dataset of the results of this study to ensure reproducibility and aid future research~\cite{Qi:TSE17}.\\
\hline
\end{tabular}}
\end{table}
}

%% file: background.tex
\section{Background \& Related Work}
\label{sec:techniques}
\blue{In this section we formally define the TCP problem, introduce our studied set of subject studied techniques, and further differentiate the novelty and research gap that our study fulfills.}

Rothermel et al.~\cite{Rothermel:01} formally defined the
test prioritization problem as finding $T'\in P(T)$, such that $\forall
T'', T''\in P(T)\wedge T''\neq T'\Rightarrow f(T')\ge f(T'')$, where $P(T)$
denotes the set of permutations of a given test suite $T$, and $f$
denotes a function from $P(T)$ to real numbers.  \blue{All of the approaches studied in this paper attempt to address the TCP problem formally enumerated above with the objective function of uncovering the highest number of faults with the smallest set of most highly prioritized test cases. As defined in previous work \cite{Henard:ICSE16, Thomas:EMSE14}, a white-box TCP approach requires access to \textit{both} the source code of subject programs, and other types of information (e.g., test code), whereas black-box techniques do not require the source code or test code of subject programs, and grey-box techniques require access to only the test-code.  Most dynamic techniques (including the ones considered in this study) are considered white-box techniques since they require access to the subject system's source code. In our study, we limit our focus to white and grey-box static TCP techniques that require only source code and test cases, and the dynamic TCP techniques that only require dynamic coverage and test cases as inputs for two main reasons:} 1) this represents fair comparison of similar techniques that leverage traditional inputs (e.g., test cases, source code and coverage info), and 2) the inputs needed by other techniques (e.g., requirements, code changes, user knowledge) are not always available in real-world subject programs. 

	\blue{In the next two subsections, we introduce the underlying methodology utilized by our studied static TCP techniques (Section~\ref{sec:static-pri}) and dynamic TCP techniques (Section~\ref{sec:dynamic-pri}). Details of our own re-implementation of these tools are discussed later in Section 3.} Additionally, we discuss existing empirical studies (Section~\ref{sec:study-pri}).
\subsection{Static TCP Techniques}
\label{sec:static-pri}

\noindent{\bf Call-Graph-Based.}
This technique builds a call graph for each test case to obtain a set of transitively invoked methods, called \textit{relevant methods}~\cite{Zhang:ICSM09}.  \blue{The test cases with a higher number of invoked methods in the corresponding call-graphs are assigned a higher test ability and thus are prioritized first. This approach is often implemented as one of two variant sub-strategies, the \textit{total} strategy prioritizes the test cases with higher test abilities earlier, and the \textit{additional} strategy prioritizes the test cases with higher test abilities excluding the methods that have already been covered by the prioritized test cases. Further research by Mei \emph{et al.} extends this work to measure the test abilities of the test cases according to the number of invoked statements as opposed to the number of invoked methods~\cite{Mei:TSE12}. The main intuition behind such an extension is that by allowing for a more granular representation of test ability (at the statement level) leads to a more effective overall prioritization scheme.}  This call-graph based technique is classified as a white box approach, whereas the other two studied static TCP techniques are grey-box approaches, requiring only test code. We consider both types of static techniques in this paper in order to thoroughly compare them to a set of techniques that require dynamic computation of coverage.

\noindent{\bf String-Distance-Based.} \blue{The key idea underlying this technique is that test cases that are textually different from one another, as measured by similarity based on string-edit distance, should be prioritized earlier \cite{Ledru:ASE12}. The intuition behind this idea is that textually dissimilar test cases have a higher probability of executing different paths within a program. This technique is a \textit{grey-box} static technique since the only information it requires is the test code. There are four major variants of this technique differentiated by the string-distance metric utilized to calculate the gap between each pair of test cases: Hamming, Levenshtein, Cartesian, and Manhattan distances.} Based on prior experimental results~\cite{Ledru:ASE12}, Manhattan distance performs best in terms of detecting faults. Thus, in our study, we implemented the string-based TCP based on the paper by Ledru \emph{et al.} \cite{Ledru:ASE12}, and chose Manhattan distance as the representative string distance computation for this technique. \blue{Explicit details regarding our implementation are given in Section \ref{sec:study}.}

\noindent{\bf Topic-Based.} \blue{This static black-box technique further abstracts the concept of using test case diversity for prioritization by utilizing semantic-level topic models to represent tests of differing functionality, and gives higher prioritization to test cases that contain different topics from those already executed \cite{Thomas:EMSE14}. The intuition behind this technique is that semantic topics, which abstract test cases' functionality, can capture more information than simple textual similarity metrics, and are robust in terms of accurately differentiating between dissimilar test cases. This technique constructs a vector based on the code of each test case, including the test case's correlation values with each semantically derived topic. It calculates the distances between these test case vectors using a Manhattan distance measure, and defines the distance between one test case and a set of test cases as the minimum distance between this test case and all test cases in the set.} During the prioritization process, the test case which is farthest from all other test cases is firstly selected and put into the (originally empty) prioritized set. Then, the technique iteratively adds the test case farthest from the prioritized set into the prioritized set until all tests have been added.

\noindent{\bf Other Approaches.} \blue{In related literature, researchers have proposed various other techniques to prioritize tests based on software requirement documents~\cite{Arafeen:ICST13} or system models~\cite{Korel:08}. Recently, Saha \emph{et al.} proposed an approach that uses software trace links between source code changes and test code derived via Information Retrieval (IR) techniques and sorts the test cases based according to the relationships inferred via the trace links, with tests more cloesly corresponding to changes being prioritized first~\cite{Saha:ICSE15}. These techniques require additional information, such as the requirement documents, system models, and code changes, which may be unavailable or challenging to collect. In this study, we center our focus on automated TCP techniques that require only the source code and the test code of subjects, including call-graph-based, string-based and topic-based techniques.}
\subsection{Dynamic TCP Techniques}
\label{sec:dynamic-pri}

\noindent{\bf Greedy Techniques.} \blue{As explained in our overview of the Call-Graph-based approach, there are typically two variants of traditional ``greedy" dynamic TCP techniques, the \textit{total} strategy and \textit{additional} strategy, that prioritize test cases based on code coverage information. The total strategy prioritizes test cases based on their absolute code coverage, whereas the additional strategy prioritizes test cases based on each test case's contribution to the total cumulative code coverage.} In our study, we implemented these techniques based on prior work by Rothermel et al. \cite{Rothermel:99}. \blue{The greedy-\textit{additional} strategy has been widely considered as one of the most effective TCP techniques in previous work~\cite{Jiang:09,zhang2013bridging}. Recently, Zhang \emph{et al.} proposed a novel approach to bridge the gap between the two greedy variants by unifying the strategies based on the fault detection probability~\cite{zhang2013bridging,Hao:TOSEM14}.}

\blue{Given that these dynamic TCP techniques utilize code coverage information as a proxy for test effectiveness, and many different coverage metrics exist, studies have examined several of these metrics in the domain of TCP including statement coverage \cite{Rothermel:99}, basic block and method coverage~\cite{Do:04}, Fault-Exposing-Potential (FEP) coverage~\cite{Elbaum:TSE02}, transition and round-trip coverage~\cite{Xu:ICST10}.}  For instance, Do \emph{et al.} use both method and basic block coverage information to prioritize test cases~\cite{Do:04}. \blue{Elbaum \emph{et al.} proposed an approach that prioritizes test cases based on their FEP and fault index coverage~\cite{Elbaum:TSE02}, where test cases exposing more potential faults will be assigned a higher priority.} 
Kapfhammer \emph{et al.} use software requirement coverage to measure the test abilities of test cases for test prioritization ~\cite{Kapfhammer:2007}.

\noindent{\bf Adaptive Random Testing.}~\blue{ Jiang \emph{et al.} were the first to apply Adaptive Random Testing~\cite{Chen:04} to TCP and proposed a novel approach, called Adaptive Random Test Case Prioritization (ART)~\cite{Jiang:09}.} ART randomly selects a set of test cases iteratively to build a candidate set, then it selects from the candidate set the test case farthest away from the prioritized set. The whole process is repeated until all test cases have been selected. \blue{As a measure of distances between test cases, ART first calculates the distance between each pair of test cases using Jaccard distance based on their coverage, and then calculates the distance between each candidate test case and the prioritized set.} \blue{Three different variants of this approach exist ({\em min}, {\em avg} and {\em max}), differentiated by the type of distance used to determine the similarity between one test case and the prioritized set.} For example, {\em min} is the minimum distance between the test case and the prioritized test case. The results from Jiang et al's evaluation illustrates that ART with {\em min} distance performs best for TCP. \blue{Thus, in our empirical study, we implemented our ART based TCP strategy following Jiang \emph{et al.}'s paper \cite{Jiang:09} and chose {\em min} distance to estimate the distance between one test case and the prioritized set.}

\noindent{\bf Search-based Techniques.} \blue{Search-based TCP techniques introduce meta-heuristic search algorithms into the TCP domain, exploring the state space of test case combinations to find the ranked list of test cases that detect faults more quickly~\cite{Li:07}.} \blue{Li \textit{et al.} have proposed two variants of search-based TCP techniques, based upon hill-climbing and genetic algorithms. The hill-climbing-based technique evaluates all neighboring test cases in a given state space, locally searching the ones that can achieve largest increase in fitness. The genetic technique utilizes an evolutionary algorithm that halts evolution when a predefined termination condition is met, e.g., the fitness function value reaches a given value or a maximal number of iterations has been reached. In our empirical study, we examine the genetic-based test prioritization approach as the representative search-based test case prioritization technique, as previous results demonstrate that genetic-based technique is more effective in detecting faults~\cite{Li:07}.}

\noindent{\bf Other Approaches.} \blue{Several other techniques that utilize dynamic program information have been proposed, but do not fit neatly into our classification system enumerated above \cite{Islam:CSMR12,Tonella:ICSM06,Nguyen:ICWS11}. Islam \emph{et al.} presented an approach that reconciles information from traceability links between system requirements and test cases and dynamic information, such as execution cost and code coverage, to prioritize test cases~\cite{Islam:CSMR12}.} Nguyen \emph{et al.} have designed an approach that uses IR techniques to recover the traceability links between change descriptions and execution traces for test cases to identify the most relevant test cases for each change description~\cite{Nguyen:ICWS11}. \blue{Unfortunately, these TCP techniques require information beyond the test code and source code (e.g., execution cost, user knowledge, code changes) which may not be available or well maintained depending on the target software project.} In this paper, we choose dynamic techniques that require only code coverage and test cases for comparison, which includes three techniques (i.e., Greedy (with total, additional strategies), ART, and Search-based). \blue{Recall that we do not aim to study the impact of coverage granularity on the effectiveness of dynamic TCPs, and opt to utilize only statement level coverage information in our experiments.  This is because previous work has established that statement-level coverage is \textit{at least as effective} as other coverage types \cite{Lu:ICSE16,Mei:TSE12}.}

\subsection{Empirical studies on TCP techniques}
\label{sec:study-pri}

	Several studies empirically evaluating TCP techniques~\cite{Kasurinen:2010, Rothermel:99, Catal:2013, Wang:2014, do06nov, Epitropakis:2015,You:2011,Smith:SAC09, Henard:ICSE16,Lu:ICSE16,Elbaum:SQI04,Elbaum:TSE02,Yoo:STVR12,Epitropakis:2015,Qu:ISSTA08} have been published.  \blue{In this subsection we discuss the details of the studies most closely related to our own in order to illustrate the novelty of our work and research gap filled by our proposed study.} Rothermel \emph{et al.} conducted a study on unordered, random, and dynamic TCP techniques (e.g., coverage based, FEP-based) applied to C programs, to evaluate their abilities of fault detection~\cite{Rothermel:99}. \blue{Elbaum \emph{et al.} conducted a study on several dynamic TCP techniques applied to C programs in order to evaluate the impact of software evolution, program type, and code granularity on the effectiveness of TCP techniques~\cite{Elbaum:TSE02}.} \blue{Thomas et. al \cite{Thomas:EMSE14} compared the topic-based TCP technique to the  static string-based, call-graph-based, and greedy-additional dynamic techniques at method-level on two subjects. However, this study is limited by a small set of subject programs, a comparison to only one dynamic technique at method-level only, and no investigation of fault detection similarity, the effects of software evolution or subject program size among the approaches.}

\Comment{Qu~\cite{Qu:ISSTA08} conducted an empirical study on seven versions of Vim to analyze the impact of software configurations on the effectiveness of several dynamic test prioritization techniques.} \blue{Do \emph{et al.} have presented a study of dynamic test prioritization techniques (e.g., random, optimal, coverage-based) on four Java programs with JUnit test suites.  This study breaks from past studies that utilize only small C programs and demonstrates that these techniques can also be effective on Java programs. However, findings from this study also  suggest that different languages and testing paradigms may lead to divergent behaviors~\cite{Do:04}. This group also conducted an empirical study to analyze the effects of time constraints on TCP techniques~\cite{Do:08}.} Henard \emph{et al.} recently conducted a study comparing white and black-box TCP techniques in which the effectiveness, similarity, efficiency, and performance degradation of several techniques was evaluated.  While this is one of the most complete studies in terms of evaluation depth, it does not consider the static techniques considered in this paper. Thus, our study is differentiated by the unique goal of understanding the relationships between purely static and dynamic TCPs.

To summarize, while each of these studies offers valuable insights, none of them provides an in-depth evaluation and analysis of the effectiveness, efficiency, and similarity of detected faults for static TCP techniques and comparison to dynamic TCP techniques on a set of mature open source software systems.  \blue{This highlights a clear research gap that exists in prior work which empirically measures the efficacy of TCP techniques.  The work conducted in this paper is meant to close this gap, and offer researchers and practitioners an extensive, rigorous evaluation of popular TCP techniques according to a varied set of metrics and experimental investigations.}

\subsection{Mutation Analysis}
\blue{Fault detection effectiveness is almost universally accepted as the measurement by which to evaluate TCP approaches \cite{Andrews:ICSE05,Just:FSE14,Lu:ICSE16}.} However, extracting a suitable set of representative real-world faults is typically prohibitively costly. \blue{Thus, researchers and developers commonly evaluate the effectiveness of TCP approaches using mutation analysis, in which a set of program variants, called mutants, are generated by seeding a large number of small syntactic errors into a seemingly ``correct" version of a program. For a given subject program, mutation operators are utilized to seed these faults (known as \textit{mutants}) into an unmodified version of the program.} It is said that a mutant is \textit{killed} by a test case when this test case is able to detect a difference between the unmodified program and the mutant. In the context of TCP research, mutation analysis is applied to a subject program to generate a large set of mutants, each containing a minor fault, and then this set is used to evaluate the effectiveness of a set of prioritized test cases.

\blue{Preliminary studies have shown mutants to be suitable for simulating real bugs in software testing experiments in controlled contexts~\cite{Andrews:05,Just:FSE14}, and mutation analysis has been used to evaluate many different types of testing approaches, including TCP techniques \cite{Elbaum:TSE02,Thomas:EMSE14,Henard:ICSE16,Lu:ICSE16,Walcott:ISST06}.} For example, Henard \emph{et al.} utilized mutation analysis to compare white-box and black-box TCP techniques \cite{Henard:ICSE16}. \blue{Lu \emph{et al.} evaluated the test case prioritization techniques in the context of evolving software systems using mutation analysis \cite{Lu:ICSE16}. Finally, Walcott \emph{et al.} proposed a time-aware test prioritization technique and evaluated their approach using mutants \cite{Walcott:ISST06}.}

\blue{Additionally, recent research has been undertaken that aims to understand the relationship between different \textit{types} of mutants (e.g., operators) and whether or not they are a suitable proxy for \textit{real} faults \cite{Kintis:APSEC10,Ammann:ICST14,Just:ICST12,Kaminski:JSS13}. Ammann \emph{et al.} proposed a framework to reduce redundant mutants and determine a minimal set of mutants for properly evaluating test cases \cite{Ammann:ICST14}. Kintis \emph{et al.} introduced several alternatives to mutation testing strategies to establish whether they adversely affect measuring test effectiveness \cite{Kintis:APSEC10}.}  \blue{However, pervious studies do not provide comments on the following in the context of TCP: 1) none of these studies has investigated the impact of the quantity of mutants utilized in TCP experiments; and 2) previous work has not examined the impact of mutants seeded according to different operators on the effectiveness of TCP approaches.} It is quite possible that TCP may perform differently when detecting different quantities or types of mutants, particularly across software projects.  Addressing these current shortcomings of past studies would allow for the verification or refutation of previous widely used experimental settings for mutation-based TCP evaluations. Thus, we aim to evaluate the effectiveness of TCP techniques in terms of detecting different quantities and types of mutants in order to understand their impact on this quality metric.

\subsection{Metrics for TCP techniques}
The Average Percentage of Faults Detected (APFD) metric is a well-accepted metric in the TCP domain \cite{Rothermel:99, Zhang:ICSM09, Rothermel:01, Do:06, Elbaum:00, Elbaum:TSE02, Elbaum:2003}, which is used to measure the effectiveness, in terms of fault detection rate, for each studied test prioritization technique. Formally speaking, let $T$ be a test suite and $T'$ be a permutation of $T$, the APFD metric for $T'$ is computed according to the following metric:
\begin{equation}\label{Equ:APFD}
    APFD=1- \frac{\sum_{i=1}^{m}TF_i}{n*m}+\frac{1}{2n}
\end{equation}
where $n$ is the number of test cases in $T$, $m$ is the number of faults, and $TF_i$ is the position of the first test case in $T'$ that detects fault $i$.

\blue{Although \apfd{} has been widely used for evaluating TCP techniques, it assumes that each test incurs the same time cost, an assumption which often doesn't hold up in practice.} Thus, Elbaum \emph{et al.} introduced another metric, called APFDc \cite{Elbaum:ICSE01}. APFDc is the cost-cognizant version of APFD, which considers both the test case execution cost and fault severity. \blue{While not as widely used as \apfd, APFDc has also been used to evaluate TCP approaches, resulting in a more detailed evaluation. \cite{Epitropakis:ISSTA15}.}  APFDc can be formally defined as follows: let $t_1, t_2, ..., t_n$ be the execution costs for all the $n$ test cases. and $f_1, f_2, ..., f_m$ be the severities of the $m$ detected faults. The APFDc metric is calculated according to the following equation:

\begin{equation}\label{Equ:APFDc}
    APFDc=\frac{\sum_{i=1}^{m}{f_i*(\sum_{j=TF_i}^{n}t_j-\frac{1}{2}t_{TF_i})}}{\sum_{i=1}^{n}t_i* \sum_{i=1}^{m}f_i}
\end{equation}

Similar to Equation \ref{Equ:APFD}, $n$ is the number of test cases in $T$, $m$ is the number of faults, and $TF_i$ is the position of the first test case in $T'$ that detects fault $i$. \blue{In our empirical study, we evaluate the performance of TCP techniques based on both of APFD and APFDc, in order to provide a complete picture of the performance of TCP techniques from the perspective of both effectiveness and efficiency.}  Additionally, we further examine the relationship between these two metrics and the resultant implications for the domain of TCP research.

%% file: study.tex
\vspace{-0.2cm}
\section{Empirical Study}
\label{sec:study}
In this section, we state our research questions, and enumerate the subject programs, test suites, study design, and implementation of studied techniques in detail.
\subsection{Research Questions (RQs):}
Our empirical study addresses the following RQs:
\begin{itemize}
\item [\textbf{RQ$_1$}] How do static TCP techniques compare with each other and with dynamic techniques in terms of \textit{effectiveness} measured by APFD?
\item [\textbf{RQ$_2$}] How do static TCP techniques compare with each other and with dynamic techniques in terms of \textit{effectiveness} measured by APFDc?
\item [\textbf{RQ$_3$}] How does the \textit{test granularity} impact the effectiveness of both the static and dynamic TCP techniques?
\item [\textbf{RQ$_4$}] \blue{How does the \textit{program size} (i.e., LOC) impact the effectiveness of both the static and dynamic TCP techniques?}
\item [\textbf{RQ$_5$}] \blue{How do static and dynamic TCP techniques perform \textit{as software evolves}?}
\item [\textbf{RQ$_6$}] How does the \textit{quantity of mutants} impact the effectiveness of  the studied TCP techniques?
\item [\textbf{RQ$_7$}] How does \textit{mutant type} impact the effectiveness of the studied TCP techniques?
\item [\textbf{RQ$_8$}] How \textit{similar} are different TCP techniques in terms of detected faults?
\item [\textbf{RQ$_9$}] How does the \textit{efficiency} of static techniques compare with one another in terms of execution time cost?
\end{itemize}
To aid in answering \textbf{RQ$_1$} and \textbf{RQ$_2$}, we introduce the following null and alternative hypotheses. The hypotheses are evaluated at the 0.05 level of significance:
\begin{description}
\item[\textbf{$H_{0}$}:] There is no statistically significant difference in the effectiveness between the studied TCPs.
\item[\textbf{$H_{a}$}:] There is a statistically significant difference in the effectiveness between the studied TCPs.
\end{description}
\input{tables/subs}

\subsection{Subject Programs, Test Suites and Faults}

We conduct our study on 58 real-world Java programs from GitHub~\cite{github}. The program names and sizes in terms of lines of code (LOC) are shown in Table~\ref{tab:sub}, where the sizes of subjects vary from 1,151 to 82,998 LoC. Our subjects are larger in size and quantity than previous work in the TCP domain~\cite{Lu:ICSE16,Henard:ICSE16,Thomas:EMSE14,Ledru:ASE12,Jiang:09}. \blue{Our methodology for collecting these subject programs is as follows. We first collect a set of 399 Java programs from GitHub that contain integrated JUnit test cases and can be compiled successfully. Then, we discarded programs which were relatively small in size (i.e., less than 1,000 LOC), or that had a very small number of test cases (i.e., less than 15 test cases at method level and five test cases at class level). Finally, we ran a set of tools to collect both the static and dynamic information (Section 3.4) and discarded programs for which the tools were not applicable. After this process we obtained our set of 58 subject programs.}

To perform this study, we checked out the most current master branch of each program, and provide the version IDs in our online appendix \cite{Qi:TSE17}. For each program, we used the original JUnit test suites for the corresponding program version. Since one of the goals of this study is to understand the impact of test granularity on the effectiveness of TCP techniques, we introduce two groups of experiments in our empirical study based on two test-case granularities: (i) the test-method and (ii) the test-class granularity. The numbers of test cases on test-method level and test-class level are shown in Columns 3 \& 4 of Table~\ref{tab:sub} respectively.

\begin{table}[t!]
\center
\caption{\label{tab:op}Muation Operators Used}
\begin{tabular}{|l|l|}
\hline
ID&Mutation Operator\\\hline\hline M0& Conditional Boundary Mutator\\
M1& Constructor Call Mutator\\ M2& Increments Mutator\\ M3& Inline
Constant Mutator\\ M4& Invert Negs Mutator\\ M5& Math Mutator\\ M6&
Negate Conditionals Mutator\\ M7& Non-Void Method Call Mutator\\ M8&
Remove Conditional Mutator\\ M9& Return Vals Mutator\\ M10& Void
Method Call Mutator\\ M11& Remove Increments Mutator\\ M12& Member
Variable Mutator\\ M13& Switch Mutator\\ M14& Argument Propagation
Mutator\\
\hline
\end{tabular}
\end{table}

One goal of this empirical study is to compare the effectiveness of different test prioritization techniques by evaluating their fault detection capabilities. Thus, each technique will be evaluated on a set of program faults introduced using mutation analysis. As mutation analysis has been widely used in regression test prioritization evaluations~\cite{zhang2013bridging,Do:05,Lu:ICSE16,Zhang:ISSTA13} and has been shown to be suitable in simulating real program faults~\cite{Andrews:05,Just:FSE14}, this is a sensible method of introducing program defects. We applied all of the 15 available mutation operators from the PIT~\cite{PIT} mutation tool (Version 1.1.7) to generate mutants for each project. All mutation operators are listed in Table~\ref{tab:op} and their detailed definitions can be found on the PIT website \cite{PIToperator} and on our online appendix~\cite{Qi:TSE17}. \blue{We utilized PIT to determine the set of faults that can be detected by the test suites for each of our subject programs.  When running the subject program's JUnit test suite via the PIT Maven plugin, test cases are automatically executed against each mutant, PIT records the corresponding test cases capable of killing each mutant.  By analyzing the PIT reports, we obtained the information (e.g., fault locations) for each mutation fault and all the test cases that can detect it. Note that the typical implementation of PIT stops executing any remaining tests against a mutant once the mutant is killed by some earlier test to save time. However, for the purpose of obtaining a set of "killable" mutants, this is undesirable. Thus, we modified PIT to force it to execute the remaining tests against a mutant even when the mutant has been killed. Since not all produced mutants can be detected/covered by test cases, only mutants that can be detected by at least one test case are included in our study.} The number of detected mutants and the number of all injected mutants are shown in Columns 5 and 6 of Table~\ref{tab:sub} respectively. As the table shows, the numbers of detected mutants range from 132 to 46,429. There are of course certain threats to validity introduced by such an analysis, namely the the potential bias introduced by the presence of equivalent and trivial mutants \cite{Andrews:TSE06,Ammann:ICST14}. We summarize the steps we take in our methodology to mitigate this threat in Section \ref{sec:threats}.

\vspace{-0.15cm}
\subsection{Design of the Empirical Study}
\vspace{-0.05cm}
As discussed previously (Section \ref{sec:techniques}), \blue{we limit the focus of this study to TCP techniques that do not require additional inputs, such as code changes or software requirements that may require extra effort or time to collect or may be unavailable}. We select two white-box and two black-box static techniques, and four white-box dynamic techniques with statement-level coverage as the subject techniques for this study, which are listed in Table~\ref{tab:techniques}. We sample from both white and black box approaches as the major goal of this study is to examine the effectiveness and trade-offs of static and dynamic TCPs under the assumption that both the source code of the subject application, as well as the test cases are available. It is worth noting that our evaluation employs \textit{two} versions of the static topic model-based technique, as when contacting the authors of \cite{Thomas:EMSE14}, they suggested that an implementation using the Mallet \cite{Mallet} tool would yield better results than their initial implementation in R \cite{Thomas:EMSE14}. There are various potential coverage granularities for dynamic techniques, such as statement-level, method-level and class-level. Previous research showed that statement-level TCP techniques perform the best~\cite{Mei:TSE12,Hao:TOSEM14}. Thus, in our study, we choose statement-level coverage for the dynamic TCP techniques. We now describe the experimental procedure utilized to answer each RQ posed above.

\input{tables/strategies}
\textbf{RQ$_1$:} The \textit{goal} of \textbf{RQ$_1$} is to compare the effectiveness of different TCP techniques, by evaluating their fault detection capabilities. Following existing work~\cite{zhang2013bridging,Lu:ICSE16}, we fixed
the number of faults for each subject program. That is, we randomly chose 500 different mutation faults and partitioned the set of all faults into groups of five (e.g., a mutant group) to simulate each faulty program version. Thus, 100 different faulty versions (i.e., 500/5 = 100) were generated for each program. If a program has less than 500 mutation faults, we use all detected mutation faults for this program and separate these faults into different groups (five faults per group).  For the static techniques, we simply applied the techniques as described in Sections \ref{sec:techniques} \& \ref{sec:impl} to the test and source code of each program to obtain the list of prioritized test cases for each mutant group. For the dynamic techniques, we obtained the coverage information of the test-cases for each program.  We then used this coverage information to implement the dynamic approaches as described in Sections \ref{sec:techniques} \& \ref{sec:impl}.  Then we are able to collect the fault detection information for each program according to the fault locations.

To measure the effectiveness in terms of rate of fault detection for each studied test prioritization technique, we utilize the well-accepted Average Percentage of Faults Detected (APFD) metric in TCP domain~\cite{Rothermel:99, Zhang:ICSM09, Rothermel:01, Do:06, Elbaum:00, Elbaum:TSE02, Elbaum:2003}. Recall that every subject program has 100 mutant groups (five mutations per group). Thus, we created 100 faulty versions for each subject (each version contains five mutations) and ran all studied techniques over these 100 faulty versions. That is, running each technique 100 times for each subject. Then, we performed statistical analysis based on the APFD results of these 100 versions. To test whether there is a statistically significant difference between the effectiveness of different techniques, we first performed an one-way ANOVA analysis on the mean APFD values for all subjects and a Tukey HSD test~\cite{Tukey}, following the evaluation procedures utilized in related work \cite{Mei:TSE12,Lu:ICSE16}. The ANOVA test illustrates whether there is a statistically significant variance between all studied techniques and the Tukey HSD test further distinguishes techniques that are significantly different from each other, as it classifies them into different groups based on their mean APFD values~\cite{Tukey}. These statistical tests give a statistically relevant overview of whether the mean APFD values for the subject programs differ significantly.  Additionally, we performed a Wilcoxon signed-rank test between each pair of TCP techniques for their average APFD value across all subject techniques, to further illustrate the relationship between individual subject programs. We choose to include this non-parametric test since we cannot make assumptions about wether or not the data under consideration is normally distributed.

\textbf{RQ$_2$:} Although \apfd{} has been widely used for TCP evaluation, it assumes that each test takes the same amount of time, which may not be always accurate in practice. The \textit{goal} of this \textbf{RQ} is to examine the effectiveness of TCP techniques in terms of the APFDc metric, which considers both the execution time and severities of detected faults. We also compare the results of the \apfdc{} with those of the APFD for understanding the performance of different types of metrics in the TCP area. However, there is no clearly-defined way to estimate the severities for the detected faults, and no widely-used tool to collect this information, making it hard to measure fault severity. Therefore, following previous work \cite{Epitropakis:ISSTA15}, we consider all faults to share the same severity level. Thus, in the context of our empirical study, APFDc reduces to the following equation:
\begin{equation}\label{Equ:APFDc}
    APFDc=\frac{\sum_{i=1}^{m}{\sum_{j=TF_i}^{n}t_j-\frac{1}{2}t_{TF_i}}}{\sum_{i=1}^{n}t_i*m}
\end{equation}
where $n$ is the number of test cases in $T$, $m$ is the number of faults, $TF_i$ is the position of the first test case in $T'$ that detects fault, and $i$, $t_1, t_2, ..., t_n$ are the execution costs for all the $n$ test cases. \blue{To measure test execution costs, we use the Maven Surefire Plugin to trace the start and end events of each test to record the corresponding execution time.} Similar as \textbf{RQ$_1$}, we performed both of an one-way ANOVA analysis on the mean AFPDc values for all subjects and a Tukey HSD test to further understand the whether there is a statistically significant variance between the performance of the studied techniques in terms of APFDc values. In addition, we further examined the relationship between the two metrics, AFPD and APFDc, to understand the differences in effectiveness of TCP techniques. We utilize the Kendall rank correlation coefficient $\tau$~\cite{sen1968estimates} to compare the results of these two metrics. Kendall rank correlation coefficient $\tau$ is commonly used to examine the relationship between two ordering quantities (i.e., observations of two variables). The coefficient ranges in value from $-1$ to $1$, with values closer to 1 indicating similarity and values closer to $-1$ indicating dissimilarity. When the value is close to $0$, these two quantities are considered independent. For example, in the context of our study, we have two quantities, APFD and APFDc values. Thus, in the context of our study, if the values of APFD values across all TCP techniques are similar to APFDc values, the Kendall tau rank coefficient $\tau$ would be closer to $1$. Otherwise, it would be closer to $-1$. Since there is no guarantee that the relationship between APFD and APFDc values are linear, we chose Kendall \taub{} coefficient in our study, following prior work \cite{Gligoric:ISSTA13}:

\begin{equation}\label{Equ:tau}\begin{split}
& \tau_b=\frac{n_c-n_d}{\sqrt{(n_0-n_1)(n_0-n_2)}}\\
& n_c=\#\mbox{ \small of concordant pairs}\\
& n_d=\#\mbox{ \small of discordant pairs}\\
& n_0=n(n-1)/2\\
& n_1=\sum_i{t_i(t_i-1)/2}\\
& n_2=\sum_j{u_j(u_j-1)/2}\\
& t_i=\#\textrm{ \small of tied values in $i$th tie group for 1st quantity}\\
& u_j=\#\mbox{ \small of tied values in $j$th tie group for 2nd quantity}\\
\end{split}
\end{equation}

\textbf{RQ$_3$} The \textit{goal} of this \textbf{RQ} is to analyze the impact of different test granularities on the effectiveness of TCP techniques. Thus, we choose two granularities: test-method and test-class levels. The test-method level treats each JUnit test method as a test case, while test-class level treats each JUnit test class as a test case.  We examine both the effectiveness and similarity of detected faults for both granularities.

\blue{\textbf{RQ$_4$} The \textit{goal} of this \textbf{RQ} is to investigate the impact of different program sizes on the effectiveness of TCP techniques. Thus, we measure the size for each subject program in terms of its Lines of Code (LOC).  To examine whether TCP technqiues tend to perform differently on programs of different sizes we classify the programs into two groups, a set of \textit{smaller} programs and a set of \textit{larger} programs. These two groups were created by ordering our subject programs in increasing order of LOC and splitting the ordered list in the middle.  This results in two groups of 29 subject programs, the first group containing smaller programs and the second group containing larger programs.}

\blue{\textbf{RQ$_5$:} The \textit{goal} of this \textbf{RQ} is to understand the effectiveness of TCP techniques in a software evolution scenario. To accomplish this we apply different TCP techniques across different versions of each subject program.  More specifically, tests are prioritized using the information from a given previous program version, and the prioritized set of test cases is then applied to faulty variants of the most recent program version. The faulty variants are created using the same methodology described for RQ$_1$.  This methodology closely follows that of previous work \cite{Lu:ICSE16} and allows us to investigate whether the performance of TCP techniques remains stable, decreases, or increases as software evolves. In our study, we collect different versions for each subject program exactly following the methodology proposed in \cite{Lu:ICSE16}. For each subject, we start from the most current version and collect one version per ten commits moving backward through the commit history. We then discard those programs that did not successfully compile and those that are not applicable to our tools. Due to the extremely large volume of data and the time cost of running these experiments, we randomly chose 12 subject programs to investigate this research question. Note that the numbers of versions (i.e., 66) and subject programs (i.e., 12) used in this work are larger than all prior TCP work considering evolutionary scenarios (e.g., the recent work by Lu et.al. ~\cite{Lu:ICSE16} used 53 versions of 8 real-world programs).}

\textbf{RQ$_6$:} The \textit{goal} of this \textbf{RQ} is to examine the impact of mutant quantity on the effectiveness for TCP techniques in terms of APFD and APFDc values. In our default experimental setting, we have 100 groups of mutation faults, and each group contains five mutants following prior work~\cite{Mei:TSE12, zhang2013bridging, Hao:TOSEM14, Lu:ICSE16}. However, in practice, the number of faults within a buggy version can be more than or less than five. Therefore, to better understand the impact of fault quantity per group, we generate different number of faults (i.e., 1 to 10) within each of the 100 constructed fault groups for each subject program. Note that we may have less than 100 fault groups when the number of mutants are small for some subjects. That is, we repeat all our prior experiments 10 times, each time recording the \apfd{} and \apfdc{} values for all studied techniques under 100 fault groups with a different number of faults (from 1 to 10). \Comment{first randomly chose 100 mutation faults to generate 100 groups of faults (each group contains one mutation fault), and evaluate the effectiveness of TCP techniques in terms of APFD and APFDc values relying on these 100 mutation groups. Then, we randomly chose another 100 mutation faults and added them into the 100 groups, and also evaluated all TCP techniques in terms of detecting these mutation groups. At this point, each group contains two mutation faults. We repeated these steps until there were ten mutation faults per group.} Finally, we perform Kendall rank \taub{} coefficient analysis to understand the relationship between the results for the mutation groups with different sizes and the results of the default setting (i.e., with 5 faults within each group). That is, we perform Kendall analysis to compare each fault-quantity setting (i.e., 100 groups mutation faults and the size of each fault group varies from 1 to 10) to the mutation faults with the default setting. Intuitively, if the values of Kendall \taub{} coefficient are close to $1$, the TCP techniques perform similarly between fault groups of varying sizes and fault groups with the default size, implying that the quantity of mutation faults does not impact TCP evaluation. 

\textbf{RQ$_7$:} The \textit{goal} of this \textbf{RQ} is to understand the impact of the mutant types (i.e., those mutants generated with different operators) on the effectiveness for TCP techniques in terms of APFD and APFDc values. Intuitively, we first classified mutants into different groups based on their corresponding operators. That is, the mutation faults generated by the same operators would be classified into the same group. In our empirical study, we utilized all 15 built-in mutation operators in PIT.\Comment{ They are {\tt negate conditional}, {\tt remove conditional}, {\tt constructor call}, non void method call, math, member variable, inline constant, increment, argument propagation, conditional boundary, switch, void method call, invert negative, return values, and remove increments.} Thus, we have 15 types of mutation faults for each subject program. We evaluate TCP techniques across these 15 types of mutation faults with the default setting, where for each operator we randomly choose 500 mutants and separate them into 100 groups (each group contains 5 mutation faults). Note that we may have less than 100 fault groups when the number of mutants are small for some mutant types. Then, TCP techniques are evaluated based on these groups of mutation faults. Finally, we compare the results for different types of mutation faults with our default fault seeding (i.e., randomly including different types of faults) under the same default setting (i.e., 100 mutated groups and each group contains 5 mutation faults). Similar as \textbf{RQ$_4$}, we chose Kendall rank tau coefficient to measure the relationship between them to check if the type of mutation fault impacts TCP evaluation.

\textbf{RQ$_8$:} The \textit{goal} of this \textbf{RQ} is to analyze the similarity of detected faults for different techniques to better understand the level of equivalency of differing strategies. It is clear that this type of analysis is important, as while popular metrics such as APFD measure the \textit{effectiveness} between two different techniques, this does not reveal the similarity of the test cases in terms of uncovered faults. For instance, let us consider two TCP techniques A and B.  If technique A achieves an APFD of $\approx 60\%$ and technique B achieves an APFD of $\approx 20\%$, while this gives a measure of relative effectiveness, the APFD does not reveal how similar or orthogonal the techniques are in terms of the faults detected.  For instance, all of the faults uncovered by top ten test cases from technique B could be different than those discovered by top ten test cases from technique A, suggesting that the techniques may be \textit{complimentary}. To evaluate the similarity between different TCP techniques, we utilize and build upon similarity analysis used in recent work \cite{Henard:ICSE16,Henard:TSE14} and construct binary vector representations of detected faults for each technique and then calculate the distance between these vectors as a similarity measure.

We employ two methodologies in order to give a comprehensive view of
the similarity of the studied TCPs.  At the core of both of these
techniques is a measure of similarity using the Jaccard distance to
determine the distance between vectorized binary representations of
detected faults (where a 1 signifies a found fault and a 0 signifies
an undiscovered fault) for different techniques across individual or
groups of subject programs. We use the following
definition \cite{Henard:ICSE16}:

\vspace{-0.2cm}
\begin{equation}\label{Equ:Jaccard}
    J(T_A^i,T_B^i)=\frac{\mid T_A^i \cap T_B^i \mid}{ \mid T_A^i \cup T_B^i \mid}
    \vspace{-0.2cm}
\end{equation}

\noindent where $T_A^i$ represents the binary vectorized discovered faults of some studied technique A after the execution of the $i^{th}$ test case in the techniques prioritized set, and $T_B^i$ represents the same meaning for some studied technique B and $0 \leq J(T_A^i, T_B^i) \leq 1$.  While we use the same similarity metric as in \cite{Henard:ICSE16}, we report two types of results: 1) results comparing the similarity of the studied static and dynamic techniques using the average Jaccard coefficient across all subjects at different test-case granularities, and 2) results comparing each technique in a pair-wise manner for each subject program. For the second type of analysis, we examine each possible pair of techniques and rank each subject program according to Jaccard coefficient as highly similar (1.0 - 0.75), similar (0.749 - 0.5), dissimilar (0.49 - 0.25), or highly dissimilar (0.249-0). This gives a more informative view of how similar two techniques might be for different subject programs. To construct both types of binary fault vectors, we use the same fault selection methodology used to calculate the APFD, that is, we randomly sample 500 faults from the set of discoverable faults for each subject.

In addition, we also want to understand whether the studied TCP techniques' most highly prioritized test cases uncover comparatively different numbers of mutants generated by different operators. Thus, for different cut points, particularly the top cut points (e.g. 10\%), we examine the both the total number and relative percentages of different types of mutants detected by each TCP technique to better understand the types of mutants which are easily detected by most highly prioritized test cases for different techniques.

\textbf{RQ$_9$:} The final \textit{goal} of our study is to understand the efficiency of static techniques, in terms of execution costs. Note that, we only focus on the efficiency of static techniques, since dynamic techniques are typically run on the previous version of a program to collect coverage information, and thus the temporal overhead is quite high and well-studied. To evaluate the efficiency of static techniques, we collect two types of time information: the time for pre-processing and the time for prioritization. The time for pre-processing contains different phases for different techniques. For example, TP$_{cg-tot}$ and TP$_{cg-add}$ need to build the call graphs for each test case. TP$_{str}$ needs to analyze the source code to extract identifiers and comments for each test case. Besides, TP$_{topic}$ needs to pre-process extracted textual information and use the R-LDA package and Mallet \cite{Mallet} to build topic models. The time for prioritization refers to the time cost for TCP on different subjects.

\vspace{-0.1cm}
\subsection{Tools and Experimental Hardware}
\vspace{-0.05cm}
\label{sec:impl}

We reimplemented all of the studied dynamic and static TCPs in Java according to the specifications and descriptions in their corresponding papers since the implementations were not available from the original authors and had to be adapted to our subjects. Three of the authors carefully reviewed and tested the code to make sure the reimplementation is reliable.

\noindent
\textbf{TP$_{cg-tot}$/TP$_{cg-add}$:} Following the paper by Zhang \emph{et al.} \cite{Zhang:ICSM09}, we use the {\em IBM T. J. Watson Libraries for Analysis} (WALA) \cite{WALA} to collect the RTA static call graph for each test, and traverse the call graphs to obtain a set of relevant methods for each test case. Then, we implement two greedy strategies (i.e., total and additional) to prioritize test cases.

\noindent
\textbf{TP$_{str}$:} Based on the paper by Ledru \emph{et al.} \cite{Ledru:ASE12}, each test case is treated as one string without any preprocessing. Thus, we directly use JDT \cite{JDT} to collect the textual test information for each JUnit test, and then calculate the Manhattan distances between test cases to select the one that is farthest from the prioritized test cases.

\noindent
\textbf{TP$_{topic-r}$ and TP$_{topic-m}$:} Following the topic-based TCP paper \cite{Thomas:EMSE14}, we first use JDT to extract identifiers and comments from each JUnit test, and then pre-process those (e.g., splitting, removing stop words, and stemming). To build topic models, we used the R-LDA package \cite{LDA} for TP$_{topic-r}$ and Mallet \cite{Mallet} for TP$_{topic-m}$. All parameters are set with previously used values \cite{Thomas:EMSE14, Chen:13}. Finally, we calculated the Manhattan distances between test cases, and selected the ones that are farthest from the prioritized test cases.

\noindent
\textbf{Dynamic TCP techniques:} We use the ASM bytecode manipulation and analysis toolset~\cite{ASM} to collect the coverage information for each test. Specifically, in our empirical study, it obtains a set of statements that can be executed by each test method or test class. The greedy techniques are replicated based on the paper by Rothermel \emph{et al.} \cite{Rothermel:99}. For the ART and search-based techniques, we follow the methodology described in their respective papers \cite{Jiang:09,Li:07}.

\noindent
\textbf{Experimental Hardware:} The experiments were carried out on Thinkpad X1 laptop with Intel Core i5-4200 2.30 GHz processor and 8 GB DDR3 RAM and eight servers with 16, 3.3 GHz Intel(R) Xeon(R) E5-4627 CPUs, and 512 GB RAM, and one server with eight Intel X5672 CPUs and 192 GB RAM. \blue{All the execution time information (i.e., both of the execution time to run TCP techniques and the execution time for each test case) was collected on the laptop to ensure that the analysis for time costs is consistent.}

\begin{figure*}[t!]
\center
\begin{subfigure}[]{0.99\textwidth}
\includegraphics[width=0.99\textwidth]{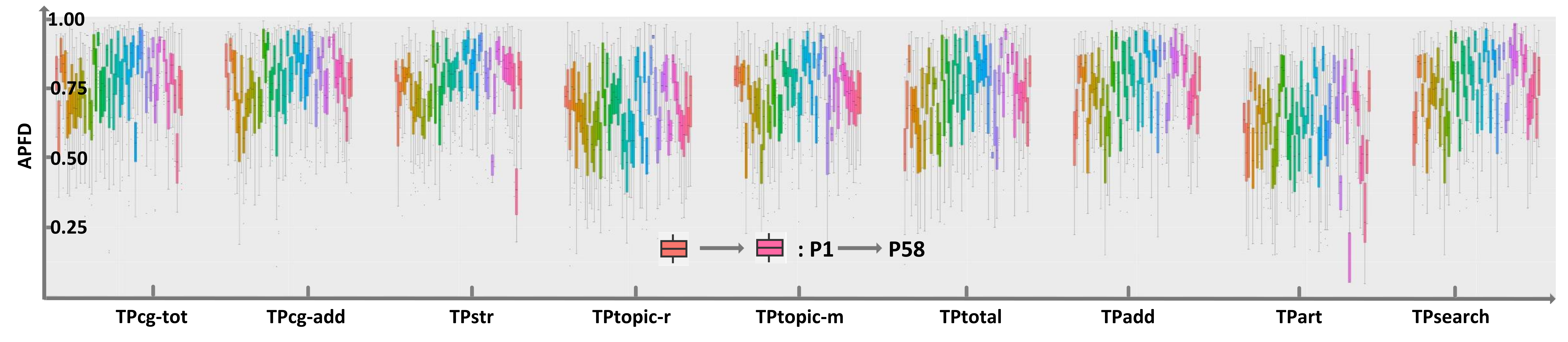}
\caption{The values of APFD on test-class level across all subject programs.\label{fig:apfd-tc}}
\end{subfigure}
\begin{subfigure}[]{0.99\textwidth}
\includegraphics[width=0.99\textwidth]{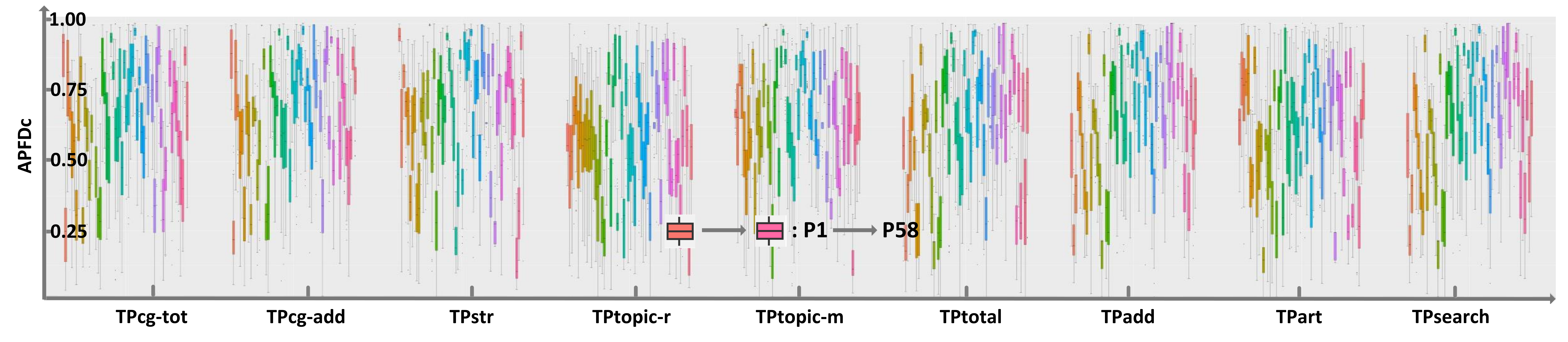}
\caption{The values of \apfdc{} on test-class level across all subject programs.\label{fig:apfdc-tc}}
\end{subfigure}
\small
\caption{The box-and-whisker plots represent the values of APFD and APFDc for different TCP techniques at test-class level. The x-axis represents the APFD and APFDc values. The y-axis represents the different techniques. The central box of each plot represents the values from the lower to upper quartile (i.e., 25 to 75 percentile).\label{fig:tc} 
}
\end{figure*}
\begin{figure*}
\center
\begin{subfigure}{0.99\textwidth}
\includegraphics[width=0.99\textwidth]{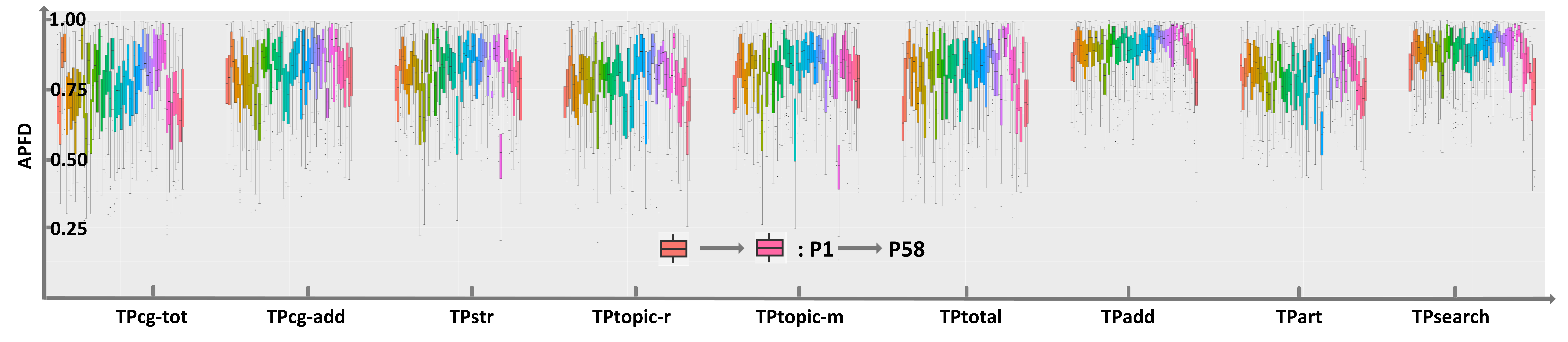}
\caption{The values of APFD on test-method level across all subject programs.\label{fig:apfd-tm}}
\end{subfigure}
\begin{subfigure}{0.99\textwidth}
\includegraphics[width=0.99\textwidth]{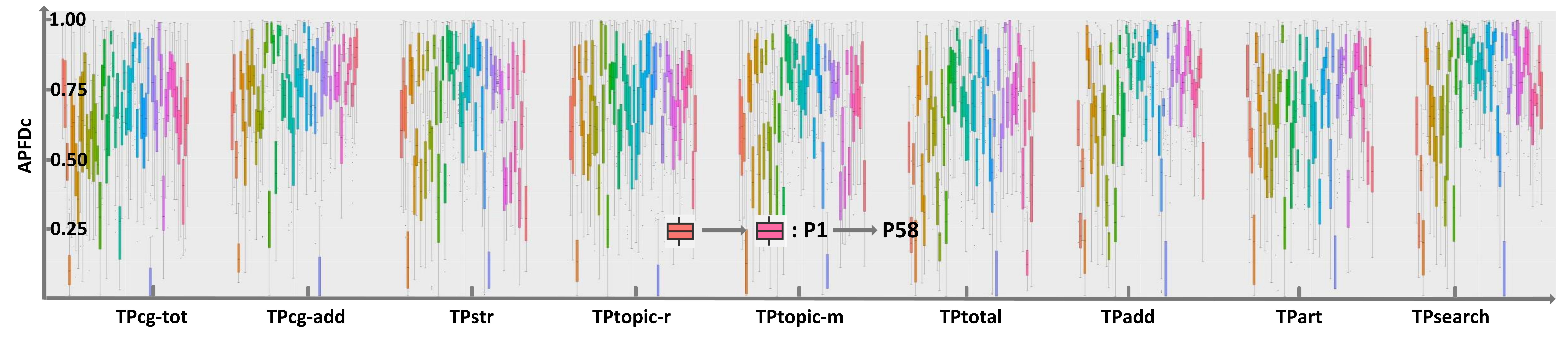}
\caption{The values of APFDc on test-method level across all subject programs.\label{fig:apfdc-tm}}
\end{subfigure}
\small
\caption{The box-and-whisker plots represent the values of APFD
    and APFDc for different TCP techniques at test-method level. The
    x-axis represents the APFD and APFDc values. The y-axis represents
    the different techniques. The central box of each plot represents
    the values from the lower to upper quartile (i.e., 25 to 75
    percentile).\label{fig:tm}}
\end{figure*}
\input{tables/statisticC.tex}
\input{tables/statistic.tex}

\begin{figure}[t]
\center

\begin{subfigure}{0.48\textwidth}
\includegraphics[width=0.98\textwidth]{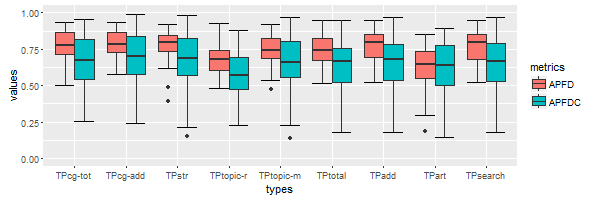}
\caption{The values of APFD and APFDc for different TCP techniques across all subject programs on test-class level.\label{fig:apfdvsapfdc-tc}}
\end{subfigure}
\begin{subfigure}{0.48\textwidth}
\includegraphics[width=0.98\textwidth]{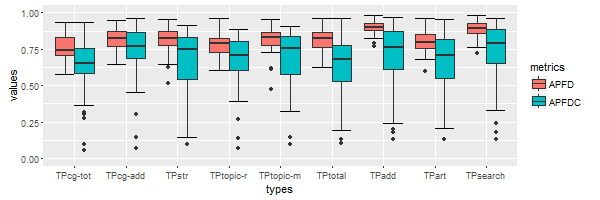}
\caption{The values of APFD and APFDc for different TCP techniques across all subject programs on test-method level.\label{fig:apfdvsapfdc-tm}}
\end{subfigure}
\small
\caption{The box-and-whisker plots represent the values of APFDc for different TCP techniques at different test granularities. The x-axis represents the APFDc values. The y-axis represents the different techniques. The central box of each plot represents the values from the lower to upper quartile (i.e., 25 to 75 percentile).\label{fig:apfdvsapfdc}}
\end{figure}

%% file: tables/subs.tex
\begin{table}
\scriptsize
\vspace{-0.2cm}
\setlength{\tabcolsep}{4.5pt}
\center\caption{\label{tab:sub}\small The stats of the subject programs: Size: \#Loc; TM: \#test cases at method level; TC: \#test cases at class level; All: \#all mutation faults; Detected: \#faults can be detected by test cases.}\vspace{-0.1cm}
\scalebox{0.95}{
\begin{tabular}{|l||c|c|c|c|c|}
\hline
&&\multicolumn{2}{c|}{Tests}&\multicolumn{2}{c|}{Mutants}\\
\cline{3-6}
Subject Programs&Size&\#TM&\#TC&{Detected}&All\\
\hline\hline
P1-geojson-jackson&1,151&44&13&301&717\\
P2-statsd-jvm-profiler&1,355&29&12&290&708\\
P3-stateless4j&1,756&61&10&392&696\\
P4-jarchivelib&1,940&22&12&655&948\\
P5-JSONassert&1,957&121&10&935&1,116\\
P6-java-faker&2,069&28&11&392&600\\
P7-jackson-datatype-joda&2,409&57&8&675&1,212\\
P8-Java-apns&3,234&87&15&412&1,122\\
P9-pusher-websocket-java&3,259&199&11&851&1,470\\
P10-gson-fire&3,421&55&14&847&1,064\\
P11-jackson-datatype-guava&3,994&91&15&313&1,832\\
P12-dictomaton&4,099&53&11&2,024&10,857\\
P13-jackson-uuid-generator&4,158&45&6&802&2,039\\
P14-JAdventure&4,416&35&10&738&5,098\\
P15-exp4j&4,617&285&9&1,365&1,563\\
P16-jumblr&4,623&103&15&610&1,192\\
P17-efflux&4,940&41&10&1,190&2,840\\
P18-metrics-core&5,027&144&28&1,656&5,265\\
P19-low-gc-membuffers&5,198&51&18&1,861&3,654\\
P20-xembly&5,319&58&16&1,190&2,546\\
P21-scribe-java&5,355&99&18&563&1,622\\
P22-jpush-api-java-client&5,462&65&10&822&2,961\\
P23-gdx-artemis&6,043&31&20&968&1,687\\
P24-protoparser&6,074&171&14&3,346&4,640\\
P25-commons-cli&6,601&317&26&2,362&2,801\\
P26-mp3agic&6,939&205&19&3,362&6,391\\
P27-webbit&7,363&131&25&1,268&3,833\\
P28-RestFixture&7,421&268&30&2,234&3,278\\
P29-LastCalc&7,707&34&13&2,814&6,635\\
P30-jackson-dataformat-csv&7,850&98&27&1,693&6,795\\
P31-skype-java-api&8,264&24&16&885&6,494\\
P32-lambdaj&8,510&252&35&3,382&4,341\\
P33-jackson-dataformat-xml&8,648&134&45&1,706&4,149\\
P34-jopt-simple&8,778&511&79&2,325&2,525\\
P35-jline2&8,783&130&16&3,523&8,368\\
P36-javapoet&9,007&246&16&3,400&4,601\\
P37-Liqp&9,139&235&58&7,962&18,608\\
P38-cassandra-reaper&9,896&40&12&1,186&5,105\\
P39-JSqlParser&10,335&313&19&15,698&32,785\\
P40-raml-java-parser&11,126&190&36&4,678&6,431\\
P41-redline-smalltalk&11,228&37&9&1,834&10,763\\
P42-user-agent-utils&11,456&62&7&376&688\\
P43-javaewah&13,293&229&11&6,307&11,939\\
P44-jsoup-learning&13,505&380&25&7,761&13,230\\
P45-wsc&13,652&16&8&1,687&17,942\\
P46-rome&13,874&443&45&4,920&10,744\\
P47-JActor&14,171&54&43&132&1,375\\
P48-RoaringBitmap&16,341&286&15&9,709&13,574\\
P49-JavaFastPFOR&17,695&42&8&46,429&64,372\\
P50-jprotobuf&21,161&48&18&1,539&10,338\\
P51-worldguard&24,457&148&12&1,127&25,940\\
P52-commons-jxpath&24,910&411&39&13,611&24,369\\
P53-commons-io&27,263&1125&92&7,630&10,365\\
P54-nodebox&32,244&293&40&7,824&36,793\\
P55-asterisk-java&39,542&220&39&3,299&17,664\\
P56-ews-java-api&46,863&130&28&2,419&31,569\\
P57-commons-lang&61,518&2388&114&25,775&32,291\\
P58-joda-time&82,998&4,026&122&20,957&28,382\\
\hline\hline
Total&714,414&15,441&1,463&245,012&542,927\\\hline
\end{tabular}
}
\vspace{-0.5cm}
\end{table}

%% file: tables/strategies.tex
\begin{table}
\vspace{-0.2cm}
\center
\setlength{\tabcolsep}{4.6pt}
\scriptsize
\caption{\label{tab:techniques}Studied TCP Techniques}
\hspace{-0.15cm}
\begin{tabular}{|l||c|c|}
\hline
Type&Tag&Description\\
\hline
\hline
\multirow{4}{*}
{Static}&$TP_{cg-tot}$&Call-graph-based (total strategy)\\
&$TP_{cg-add}$&Call-graph-based  (additional strategy)\\
&$TP_{str}$&The string-distance-based\\
&$TP_{topic-r}$&Topic-model-based using R-lda package\\
&$TP_{topic-m}$&Topic-model-based using Mallet\\
\hline
\hline
\multirow{4}{*}
{Dynamic}&$TP_{total}$&Greedy total  (statement-level)\\
&$TP_{add}$&Greedy additional (statement-level)\\
&$TP_{art}$&Adaptive random (statement-level)\\
&$TP_{search}$&Search-based (statement-level)\\
\hline
\end{tabular}
\vspace{-0.6cm}
\end{table}

%% file: tables/statisticC.tex
\begin{table*}[t]
\center
\scriptsize
\caption{\small Results for the ANOVA and Tukey HSD tests on the average APFD and APFDc values at test-class level, which are depicted in Figure \ref{fig:tc}. The last column shows the results for Kendall tau Rank Correlation Coefficient $\tau_{b}$ between the average APFDc and average APFD.\label{tab:statC}}
\begin{tabular}{|c|c|c|c|c|c|c|c|c|c|c|c|c|}
\hline
Metrics&Analysis&{TP$_{cg-tot}$}&{TP$_{cg-add}$}&{TP$_{str}$}&{TP$_{topic-r}$}&{TP$_{topic-m}$}&{TP$_{total}$}&{TP$_{add}$}&{TP$_{art}$}&{TP$_{search}$}&p-value&\taub\\
\hline\hline
\multirow{2}{*}{APFD}&Avg&0.778&0.790&0.777&0.675&0.745&0.738&0.769&0.633&0.765&\multirow{2}{*}{1.777e-18}&\multirow{4}{*}{0.722}\\\cline{2-11}
&HSD&A&A&A&B&A&A&A&B&A&&\\\cline{1-12}\cline{1-12}
\multirow{2}{*}{APFDc}&Avg&0.652&0.679&0.667&0.574&0.657&0.614&0.650&0.612&0.649&\multirow{2}{*}{0.154}&\\\cline{2-11}
&HSD&A&A&A&A&A&A&A&A&A&&\\\hline
\end{tabular}
\end{table*}

%% file: tables/statistic.tex
\begin{table*}[t]
\center
\scriptsize
\caption{\small Results for the ANOVA, and Tukey HSD tests on the average APFD and APFDc values at test-method level, which are depicted in Figure \ref{fig:tm}. The last column shows the results for Kendall tau Rank Correlation Coefficient $\tau_{b}$ between the average APFDc and average APFD.\label{tab:stat}}
\begin{tabular}{|c|c|c|c|c|c|c|c|c|c|c|c|c|c|}
\hline
Metrics&Analysis&{TP$_{cg-tot}$}&{TP$_{cg-add}$}&{TP$_{str}$}&{TP$_{topic-r}$}&{TP$_{topic-m}$}&{TP$_{total}$}
&{TP$_{add}$}&{TP$_{art}$}&{TP$_{search}$}&p-value&$\tau_{b}$\\
\hline\hline
\multirow{2}{*}{APFD}&Avg&0.764&0.818&0.813&0.781&0.817&0.809&0.898&0.798&0.885&\multirow{2}{*}{2.568e-28}&\multirow{4}{*}{0.556}\\\cline{2-11}
&HSD&C&B&B&BC&B&B&A&BC&A&&\\\cline{1-12}\cline{1-12}
\multirow{2}{*}{APFDc}&Avg&0.638&0.737&0.671&0.678&0.679&0.633&0.708&0.669&0.735&\multirow{2}{*}{0.053}&\\\cline{2-11}
&HSD&A&A&A&A&A&A&A&A&A&&\\\hline
\end{tabular}
\end{table*}

%% file: result.tex
\section{Results}
\label{sec:res}
In this section, we outline the experimental results to answer the \textbf{RQs} listed in Section~\ref{sec:study}.

\subsection{RQ$_1$ \& RQ$_2$ \& RQ$_3$: Effectiveness of Studied Techniques Measured by APFD and APFDc at Different Granularities}

\Comment{\QI{update figure1, 2 table 3, 4 to put test class level
    results together (including apfd and apfdc), and put test method
    level results together}}

\subsubsection{Results at Test Class Level}
\label{subsubsec:test-class}

The values of APFD across all subjects at the test class level are shown in Figure~\ref{fig:apfd-tc} and Table \ref{tab:statC}. Based on the results, we observe that, somewhat surprisingly at the test-class level, the static TP$_{cg-add}$ technique performs the best across all studied TCP techniques (including all dynamic techniques) with an average APFD value of 0.790 (see Table~\ref{tab:statC}). Among the static techniques, TP$_{cg-add}$ performs best, followed by TP$_{cg-tot}$, TP$_{str}$, TP$_{topic-m}$ and TP$_{topic-r}$.  The best performing dynamic technique at class-level is TP$_{add}$ followed by TP$_{search}$, TP$_{total}$, and TP$_{art}$.  It is notable that at test-class level granularity, the most effective static technique TP$_{cg-add}$ performs even better than the most effective dynamic technique TP$_{add}$ in terms of \apfd, i.e., 0.790 versus 0.769. The experimental results on \apfdc{} values further confirm the above finding. Shown in Figure~\ref{fig:apfdc-tc} and Table~\ref{tab:statC}, the static TP$_{cg-add}$ technique outperforms all the studied TCP techniques with an average \apfdc{} value of 0.679, whereas even the most effective dynamic TP$_{add}$ only achieves an average \apfdc{} value of 0.650. Furthermore, the Kendall \taub{} Rank Correlation value of 0.722 also demonstrates that \apfdc{} values are generally consistent with \apfd{} values at the test class level. Therefore, at the test-class level, the call-graph based strategies can even outperform dynamic-coverage based strategies, which is notable. Additionally, overall the static techniques outperform the dynamic techniques at the test-class level. One potential reason for this is that many program statements are covered several times by tests at the test-class level, making the traditional dynamic techniques less precise, since they do not consider the number of times that a statement is covered.

While Figure~\ref{fig:tc} shows the detailed APFD and APFDc values for each studied subject at test-class level, Figure~\ref{fig:apfdvsapfdc-tc} further shows the ranges of APFD and \apfdc{} values across all subjects at test-class level, reflecting the robustness of the studied approaches across both metrics. \Comment{\KEVIN{I think we should move this figure closer to where it is discussed, right now it is a few pages away which makes for more difficult reading}. \QI{It is one page away, and there is no space for this figure in current page. I think it is okay.}} For \apfd{}, the range of average values across all subjects at test-class level for TP$_{add}$ is the smallest (i.e.,0.523-0.947), implying that the performance of TP$_{add}$ is usually stable despite differing subjects for this metric. Conversely, the ranges of APFD values for TP$_{str}$ and TP$_{art}$ are much larger (0.391-0.917 for TP$_{str}$, 0.187-0.852 for TP$_{art}$), implying that their performance varies across different types of subjects. However, we observe different trends for the APFDc metric\Comment{ (see Fig \ref{fig:apfdvsapfdc-tc})}. The ranges of \apfdc{} values are all much larger than those of \apfd{} values. This is most likely due to the fact that \apfdc{} considers execution times, which we found to be randomly distributed, resulting in a larger variation in results across different subjects.

To further investigate the finding that static techniques tend to have a higher variance in terms of effectiveness depending on the program type, we investigated further by inspecting several subject programs. One illustrative example is that {\em scribe-java} scores $0.646$ and $0.606$ for the average values of APFD under TP$_{str}$ and TP$_{topic-r}$ respectively, which are notably worse than the results of TP$_{cg-tot}$ (0.718) and TP$_{cg-add}$ (0.733). To understand the reason for this discrepancy, we analyzed the test code and found that {\em Scribe-java} is documented/written more poorly than other programs. For instance, the program uses meaningless comments and variable names such as `$param1$', `$param2$', `$v1$', `$v2$' etc.  This confirms the previously held notion \cite{Thomas:EMSE14} that static techniques which aim to prioritize test-cases through text-based diversity metrics experience performance degradation when applied to test cases written in a poor/generic fashion. \blue{It also suggests that researchers may take the subject characteristics into account when choosing TCP techniques in future work.}

\input{tables/W-test.tex}
\input{tables/W-test-APFDC.tex}

\subsubsection{Results at Test Method Level}
\label{subsubsec:test-method}

To further answer \textbf{RQ$_3$} we ran all of the subject TCP techniques on the subject programs at the test-method level so that we can compare to the results at the test-class level outlined above (see Section \ref{subsubsec:test-class}). The results are shown in Figure~\ref{fig:tm} and Table~\ref{tab:stat}. In terms of \apfd, when examining the static techniques with the \textit{test-method granularity}, they perform differently as compared to the results on the test-class level. For example, although TP$_{cg-add}$ still performs the best among static techniques, it is inferior to the most effective dynamic technique TP$_{add}$ (0.818 versus 0.898). This finding is consistent with previous studies \cite{Hao:TOSEM14}. Also, surprisingly, TP$_{topic-m}$ (0.817) achieves almost the same average \apfd{} values as TP$_{cg-add}$, followed by TP$_{str}$, TP$_{topic-r}$ and TP$_{cg-tot}$ respectively. It is worth noting that the effectiveness of the topic-model based technique varies quite dramatically depending on the tools used for its implementation: Mallet \cite{Mallet} significantly outperforms the R-based implementation. Also, there is less variation in the \apfd{} values at the test-method level compared to those at the test-class level, as shown in Figure \ref{fig:tm} and Figure \ref{fig:apfdvsapfdc}. 

In terms of \apfdc{}, the results for test-method level are generally consistent with the results on test-class level. For example, while TP$_{search}$ tend to be the most effective dynamic technique, the static TP$_{cg-add}$ outperforms all the studied static and dynamic techniques. The likely reason is that dynamic techniques tend to favor tests with higher coverage, which tend to cost more time to execute, leading to limited effectiveness in actual time cost reduction. The results of the HSD analysis on the APFDc values at the test-method level, indicate that all techniques are grouped into the same level (level A), implying that different TCP techniques share similar performance based on APFDc values, which is also consistent with the results of \apfdc{} values at the test-class level. When examining the ranges of APFDc values for the test-method level (see Fig. \ref{fig:tm} and Fig. \ref{fig:apfdvsapfdc}), we find the APFDc values vary dramatically between subject programs. When comparing the results of APFD and APFDc values at the test-method level, the Kendall \taub{} rank coefficient \taub{} is 0.556, impling that the APFDc results are less consistent with the \apfd{} results at the test-method level. The reason is likely that test execution time distributions which are uncontrolled have large impacts the more effective/stable test-method-level results.

In addition, as a whole, the effectiveness of the dynamic techniques outpaces that of the static techniques at method-level granularity for the APFD metric, with TP$_{add}$ performing the best of all studied techniques (0.898). For the cost-cognizant APFDc metric, although there are no clear trends, the static techniques tend to perform even better than dynamic techniques, indicating the limitations of dynamic information for actual regression testing time reduction.  Overall, on average, almost all static and dynamic TCPs perform better on the test-method level as compared to the results on the test-class level in terms of both \apfd{} and \apfdc{}.  Logically, this is not surprising, as using a
finer level of granularity (e.g., prioritizing individual test-methods) gives each technique more flexibility, which leads to more accurate targeting and prioritization.

Finally, to check for statistically significant variations in the mean APFD and \apfdc{} values across all subjects and confirm/deny our null hypothesis for \textbf{RQ$_1$} and \textbf{RQ$_2$}, we examine the results of the one-way ANOVA and Tukey HSD tests. The ANOVA test for \apfd{} values, given in the second to last column of Tables \ref{tab:statC} \& \ref{tab:stat}, are both well below our established significance threshold of 0.05, thus signifying that the subject programs are statistically different from one another.  This rejects the null hypothesis $H_0$ and we conclude that there are statistically significant differences between different TCP techniques in terms of \apfd.  The results of the Tukey HSD test also illustrate the statistically significant differences between the static and dynamic techniques by grouping the techniques into categories with \textit{A} representing the best performance and the following letters (e.g., \textit{B}) representing groups with worse performance.  We see that the groupings are similar for static and dynamic techniques.  In order to illustrate the individual relationships between strategies, we present the results of the Wilcoxon signed rank test for all pairs of techniques at both granularity levels in Tables \ref{tab:stat-w} and~\ref{tab:statC-w}.  \blue{The shaded cells represent statistically significant differences between techniques across all the subjects (e.g., $p < 0.05$). The Wilcoxon signed rank test further confirms that different techniques have statistically different \apfd{} values at both test-class and test-method levels, as indicated by the shaded boxes. On the contrary, the results for \apfdc{} ANOVA and HSD tests lead to different observations -- different techniques \textit{generally} do not have statistically different \apfdc{} values (as shown in Tables \ref{tab:statC} \& \ref{tab:stat}), indicating that both static and dynamic techniques tend to perform similalrly for \apfdc{} values. The Wilcoxon signed rank test for \apfdc{} values of all pairs of techniques is shown in Table~\ref{tab:statC-w}. The small number of shaded cells (i.e., $p < 0.05$) further confirms that different techniques tend to perform equivalently for \apfdc.} The likely reason for this is that \apfdc{} is impacted by an additional randomly-distributed factor, i.e., tests tend to have randomly distributed execution times, leading to the observed results.  It should be noted that in contrast to our previous work \cite{Luo:FSE16}, our results for the HSD show less variance between the different approaches for APFD at both test-class and test-method level.  This means that the approaches were grouped in fewer differing groups by the HSD test, indicating performance that is more comparatively similar.  This illustrates the affect of generalizing across more subject programs.

\Comment{For test method level, it is clear that the dynamic techniques
outperform the static, as far more dynamic techniques are ranked in
the better performing categories.}

In summary we answer \textbf{RQ$_1$}, \textbf{RQ$_2$} \& \textbf{RQ$_3$} as follows:

\begin{framed}
{\bf RQ$_1$:} There is a statistically significant difference between the \apfd{} values of the two types (e.g., static and dynamic) of studied techniques. On average, static technique TP$_{cg-add}$ is the most effective technique at test-class level, whereas dynamic technique TP$_{add}$ is the most effective technique at test-method level. Overall, the static techniques outperform the dynamic ones at test-class level, but the dynamic techniques outperform the static ones at test-method level.
\end{framed}
\vspace{-0.2cm}
\begin{framed}{\vspace{-0.1cm} {\bf RQ$_2$:} For the APFDc values, there is no statistically significant difference between the studied static and dynamic techniques. APFDc values are generally consistent with APFD values at test-class level but relatively less consistent at test-method level. Similar to the results from RQ$_1$, on average, static TP$_{cg-add}$ technique is the most effective technique at the test-class level, and the static techniques outperform the dynamic ones as a whole at test-class level. However, at test-method level, TP$_{cg-add}$ also performs best overall, indicating the superiority of static techniques to dynamic techniques in actual regression testing time reduction. Additionally, APFDc values vary more dramatically across all subject programs compared to AFPD. \Comment{Overall for the method level, there is no general trend of one type of technique outperforming the other.}
}
\end{framed}

\begin{framed}{
{\bf RQ$_3$: }The test granularity significantly impacts the effectiveness of TCP techniques in terms of both \apfd{} and \apfdc{}, although the \apfdc{} metric is affected to a much lesser extent. All the studied techniques perform better at test-method level as compared to test-class level. There is also less variation in the APFD values at method-level as compared to class-level, which signifies that the performance as measured by this metric is more stable at test-method level across the studied techniques.}
\end{framed}

\input{tables/statisticSmallC.tex}
\input{tables/statisticLargeC.tex}
\input{tables/statisticSmall.tex}
\input{tables/statisticLarge.tex}

\blue{
\subsection{Impact of Subject Program's Size}
Since developers may apply TCP techniques to subject systems in various sizes in practice, it is important to understand the potential impact of program size on the performance of TCP techniques. Thus we examine the differences between the performance of our studied TCP techniques on our 29 smaller subject systems and 29 larger subject systems. Table~\ref{tab:statSmallC} and Table~\ref{tab:statLargeC} present the TCP results at the test-class level on smaller and larger subject systems, respectively; similarly, Table~\ref{tab:statSmall} and Table~\ref{tab:statLarge} present the TCP results at the test-method level on smaller and larger subject systems, respectively.}

\blue{From the tables, we can make the following observations. First, TCP techniques tend to perform better on larger subject systems than smaller subject systems. For example, for both test-class and test-method level, all the studied TCP techniques perform better on larger subject systems in terms of both APFD and \apfdc. One potential reason is that larger subject systems tend to have more tests, leaving enough room for TCP techniques to reach optimization thresholds. This finding also demonstrates the scalability of the studied TCP techniques. Second, at the test-class level, static TCP techniques tend to outperform dynamic TCP techniques in terms of both \apfd{} and \apfdc{} on both subsets of subject systems; in contrast, at the test-method level, static TCP techniques are inferior to dynamic TCP techniques on both subsets of subjects in terms of \apfd, while TP$_{cg-add}$ outperforms all the other studied dynamic and static TCP techniques on both subsets of subjects in terms of \apfdc. This finding is consistent with our findings in {\bf RQ$_1$} and {\bf  RQ$_2$}, indicating that subject size does not impact our findings when comparing the relative performance of the studied TCP techniques according to \apfd{} and \apfdc{}. Third, most studied TCP techniques perform better at test-method level as compared to test-class level in terms of both \apfd{} and \apfdc{} on both subsets of our subjects. This observation is also consistent with our comparative findings for {\bf RQ$_3$}.
\begin{framed}{\vspace{-0.1cm}
{\bf RQ$_4$: }All the studied TCP techniques tend to perform better on larger subject systems, indicating the scalability of the studied TCP techniques. However, when comparing the performance of different TCP techniques to each other on either the large or small programs, we find results consistent to using the entire set of programs (both in terms of APFD(c) and differing test-case granularities). Thus we can conclude that program size has little effect when comparing the relative performance of TCP techniques on a given subject. \vspace{-0.1cm}}
\end{framed}
\vspace{-0.2cm}
}

\subsection{Impact of Software Evolution}

\input{evolution-Class}
\input{evolution-Method}

\blue{
Figure~\ref{fig:ev-Class} and Figure~\ref{fig:ev-Method} present the impact of software evolution on the studied TCP techniques at the test-class and test-method levels, respectively. In each figure, each row presents both the \apfd{} and \apfdc{} results on the corresponding subject. In each sub-figure, the x-axis presents the versions used as the old version during software evolution (note that the most recent versions are always used as the new version during software evolution), while the y-axis presents the \apfd{} or \apfdc{} values. We show the \apfd{} or \apfdc{} distributions of different technique using box-plots of different colors, where the boxes represent the 25th to 75th percentiles, the centerlines represent the median values, and the dots represent the outlier points. Due to the limited space, we only show the results of two subject programs. The results of all twelve subject programs can be found in our online appendix \cite{Qi:TSE17}. Following prior work~\cite{Lu:ICSE16}, using different versions as the old version during software evolution allows us to understand the impact of software evolution on TCP in details.  To illustrate, for a project with $n$ versions, where ($n>$2), we will have a set of $n-1$ results for applying each studied TCP technique. That is, running TCPs on older program versions and then applying the prioritized set of test cases on the faulty variants of the most recent version (i.e., the latest versions with mutants) allows us to understand the performance of TCP techniques in an evolutionary scenario. Note that more recent project versions may have tests not included in older project versions; we ignore such tests since the studied techniques would not be able to prioritize those tests based on old project versions.} 

\blue{If software evolution impacts TCP effectiveness, using earlier program versions for test prioritization would likely be less effective than using more up-to-date versions for test prioritization due to code changes. In other words, \apfd/\apfdc{} values should increase when using newer versions for prioritization. However, we observe no such trend for either \apfd{} or \apfdc{} for any TCP technique on any studied subject at the test-class or test-method level. This observation confirms prior work~\cite{Lu:ICSE16, Henard:ICSE16} that code changes do not impact the effectiveness of dynamic TCP techniques in terms of \apfd. Furthermore, our work is the first to illustrate that the same finding holds for static TCP techniques as well as the more practical \apfdc{} metric. These results most likely arise due to the fact that all studied TCP techniques approximate fault detection capabilities based on a certain set of criteria (such as call graphs, textual information, or code coverage), and software evolution usually does not result in large relative changes between commits for these different criteria (e.g., some tests may always have higher code coverage throughout project evolution).}

\blue{We also find that the performance comparison in terms of APFD between dynamic and static TCP techniques is not impacted by software evolution. For instance, static TCP techniques tend to outperform dynamic TCP techniques in terms of both \apfd{} at the test-class granularity on most subjects, while dynamic TCP techniques tend to outperform static TCP techniques in terms of \apfd{} at the test-method granularity on most subjects. \apfdc{} values tend to exhibit more variance during software evolution. For example, for the {\tt javapoet} subject at test-class level, the static TP$_{str}$ technique outperforms all other  techniques when using $V_8$ information to prioritize tests for $V_8$, while the dynamic TP$_{art}$ technique performs the best when using $V_1$ information to prioritize tests for $V_8$. One potential explanation for this observation is that tests with similar fault detection capabilities may have totally different execution times during evolution, causing high variances between APFDc values. }
\vspace{-0.2cm}
\blue{\begin{framed}{\vspace{-0.1cm}
      {\bf RQ$_5$: } On average, software evolution does not have a clear impact on the measured effectiveness of the studied TCP techniques. Corroborating results of RQ$_1$ and RQ$_2$, we find that the APFD values for techniques tend to exhibit lower variance than APFDc values.}
\end{framed}
}

\subsection{Impact of Mutant Quantities on TCP Effectiveness}

\input{tables/APFD-sizes.tex}
\input{tables/APFDC-sizes.tex}

Prior works examining TCP techniques generally directly seed a certain number of faults to form a faulty version (or groups of faulty versions) to investigate TCP effectiveness according to the APFD or APFDc, similar to the setup used in our study to answer RQ$_1$-RQ$_3$. However, we wish to further analyze the impacts of the quantity of mutants utilized in experimental settings and whether or not this impacts the effectiveness of techniques. The experimental results for \apfd{} and \apfdc{} are shown Tables \ref{tab:APFD-sizes} and \ref{tab:APFDC-sizes}, respectively. In each table, Column 1 lists the test-case granularities studied, column 2 lists the number of mutants seeded into each faulty version/group, columns 3-11 present the \apfd{}/\apfdc{} results for each studied technique, and finally the last column presents the Kendall \taub{} Rank Correlation Coefficient between the average values with each fault quantity and our default settings (shown in Tables \ref{tab:statC} and \ref{tab:stat}). From the tables, we make the following observations. For both \apfd{} and \apfdc{} values, the mutant quantity does not dramatically impact the results for all of the studied techniques. For example, at the test-class level, the average \apfd{} values of TP$_{cg-add}$ range from 0.786 to 0.790 for all the studied\ fault quantity settings, while its \apfdc{} values range from 0.678 to 0.681. This finding indicates that the effectiveness of the studied techniques when seeding any number of mutants into each group will be roughly equivalent, demonstrating the validity of the mutant seeding processes of prior TCP work~\cite{Lu:ICSE16, Mei:TSE12, zhang2013bridging, Hao:TOSEM14}. The largest impact that fault quantities had were for the \apfdc{} metric at the test-method level. The likely reason for this is that the test-method level techniques prioritize tests at a finer granularity, and thus are more sensitive to the impact of execution time. For example, \apfdc{} of fault groups with only one fault in each group only considers the time to detect only the first fault (while \apfdc{} of fault groups with 5 faults in each group considers the time to detect all the 5 faults), leading to the higher variance.

\begin{framed}{\vspace{-0.1cm}
{\bf RQ$_6$: }The quantity of mutants used, as stipulated in the experimental settings of mutation analysis-based evaluations of TCP approaches, does not significantly impact the effectiveness of TCP techniques in terms of either \apfd{} or \apfdc{}, demonstrating the validity of the fault seeding process of prior work in this context.\vspace{-0.1cm}}
\end{framed}
\vspace{-0.2cm}

\input{tables/APFD-types.tex}
\input{tables/APFDC-types.tex}

\subsection{Impact of Mutant Types on TCP Effectiveness} To answer RQ$_7$, we further investigate whether different mutant types may impact TCP results in terms of either \apfd{} and \apfdc{}. The experimental results for \apfd{} and \apfdc{} are shown Tables \ref{tab:APFD-types} and \ref{tab:APFDC-types}, respectively. In each table, column 1 lists the test-case granularities studied, column 2 lists the different types of mutants seeded into each faulty version, columns 3-11 present the \apfd{}/\apfdc{} results for each studied technique, and finally the last column presents the Kendall \taub{} Rank Correlation Coefficient between the average \apfd{}/\apfdc{} values with each fault type and our default settings (shown in Tables \ref{tab:statC} and \ref{tab:stat}). 

	From the tables, we can make the following observations. \blue{First, overall the vast majority of the studied mutant types tend to have a medium to high coefficient (i.e., the range of Kendall correlation coefficient values is from 0.5 to 1.0). This implies that the performance of TCP techniques when applied to detecting only certain mutant types highly correlates to the performance observed when applied to detecting all mutants.} This indicates that the findings in prior work on TCP (including this work) generally hold across mutants seeded with differing mutation operators. 
\\

	\blue{Second, we observe that there are several mutant types with low correlation with our default fault seeding, e.g., the {\tt Invert Negs Mutator} and the \texttt{Switch Mutator} have the lowest correlation in both studied test granularities for both \apfd{} and \apfdc{}. Upon further investigation, we found one likely explanation to be the small number of mutants generated by such mutators. For example, the number of {\tt Invert Negs Mutator} is quite small as compared to other type of mutation faults (since it is only applicable to the cases of negative numbers), thus the results are dramatically different as compared to the results of mutation faults with the default setting. The \texttt{Switch Mutator} also has small Kendall correlation coefficient values as compared to other mutators. This is due to the fact that like the \texttt{Invert Negs Mutator} -- the number of \texttt{Switch Mutator} faults is quite small as compared to other type of mutation faults (since it is only applicable to the cases of switches, which are not intensively used in common programs). Thus, the results are dramatically different as compared to the results of mutation faults with the default setting.} Furthermore, we also observe that some mutation faults are more subtle than others.  For example, the mutation faults created by {\tt Invert Negs Mutator} tend to be more subtle than other types of mutation faults. For example, the {\tt Invert Negs Mutator} operator simply inverts negation of an integer and floating point number (e.g., changing ``{\tt return -i;}'' into ``{\tt return i;}''), while {\tt Non-Void Method Call Mutator} or {\tt Void Method Call Mutator} directly removes an entire method invocation. The subtle mutation faults introduced by {\tt Invert Negs Mutator} can be harder to detect, making various static and dynamic techniques perform worse on those faults since the coverage or call graph information won't provide precise guidance. 

\begin{framed}{
{\bf RQ$_7$: } The mutation operators used to seed faults, as stipulated in the experimental settings of mutation analysis-based evaluations of TCP approaches, do not significantly impact the effectiveness of TCP techniques in terms of either \apfd{} or \apfdc{}, demonstrating the comparative validity of the fault seeding process of prior work in this context.\vspace{-0.1cm}}
\end{framed}

\input{tables/JD-10percent.tex}
\input{tables/JD-50percent.tex}
\begin{figure}[tb]
\centering
\begin{subfigure}{0.49\textwidth}
\includegraphics[width=1.01\columnwidth]{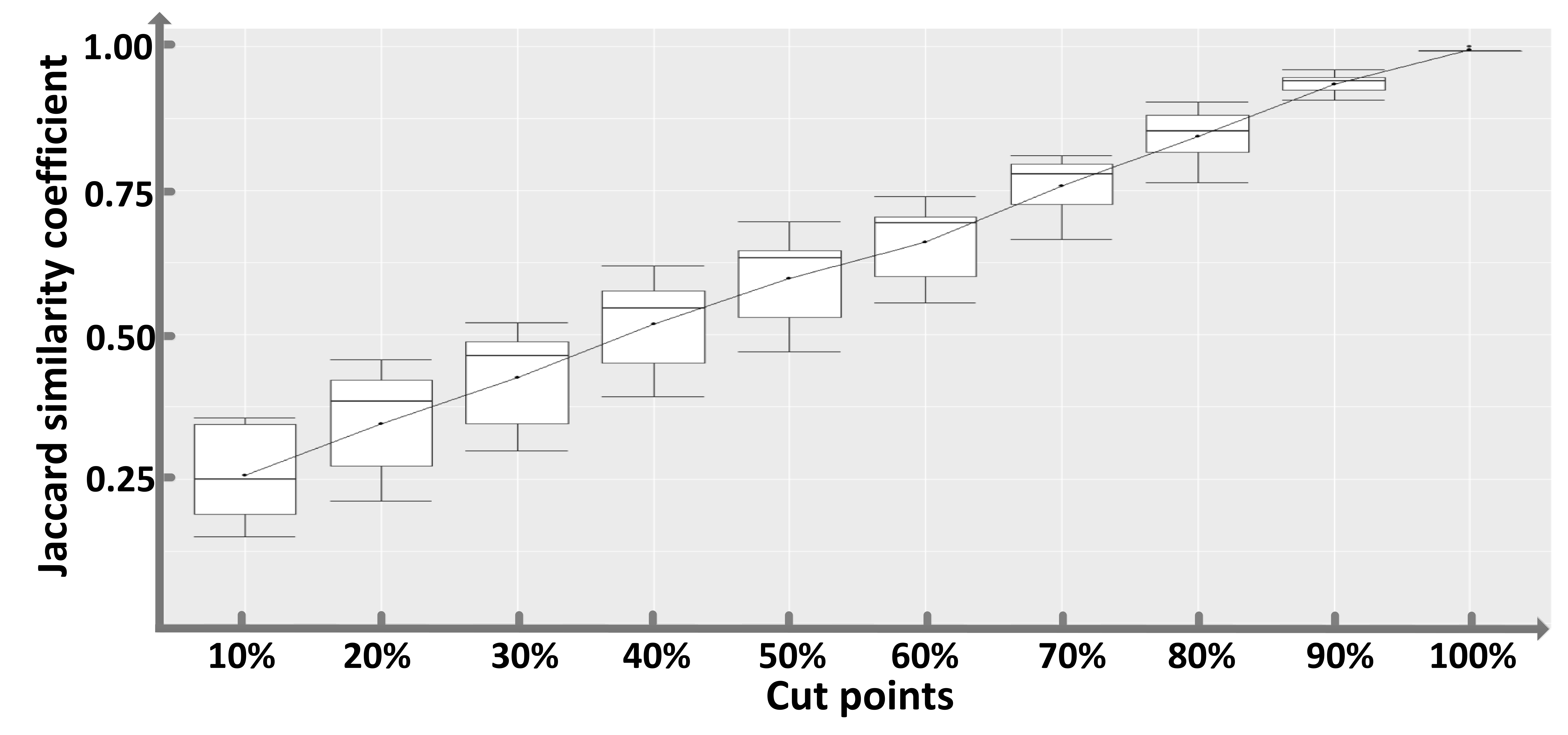}\label{fig:jac-c}
\caption{Class-level results}
\end{subfigure}
\begin{subfigure}{0.49\textwidth}
\includegraphics[width=1.01\columnwidth]{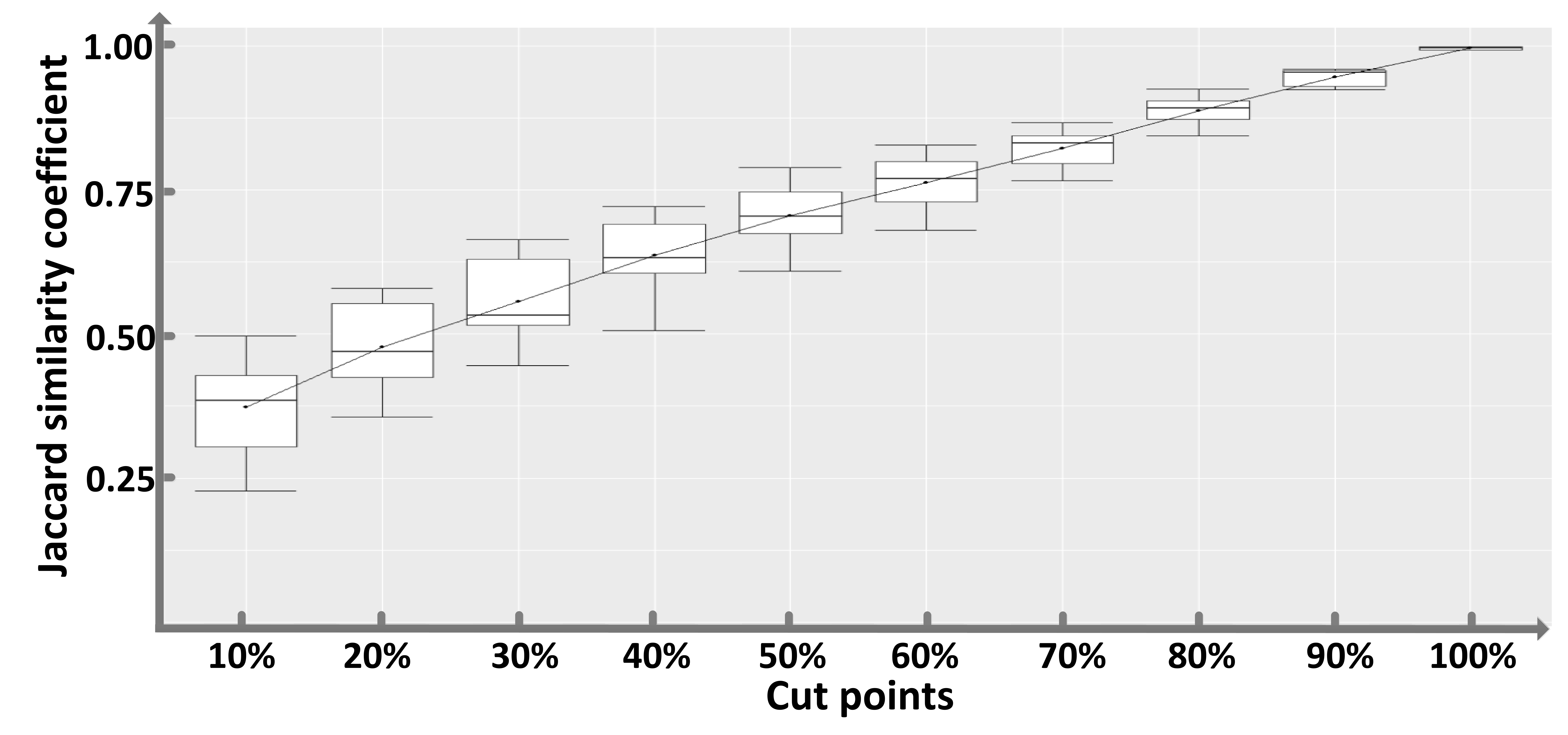}\label{fig:jac-m}
  \caption{Method-level Results}
\end{subfigure}
  \caption{\label{fig:jac}Average Jaccard similarity of faults detected between static and dynamic techniques across all subjects at method and class-level granularity. }
\end{figure}

\begin{figure*}[t]
\centering
\begin{subfigure}{0.49\textwidth}
\includegraphics[width=0.98\columnwidth]{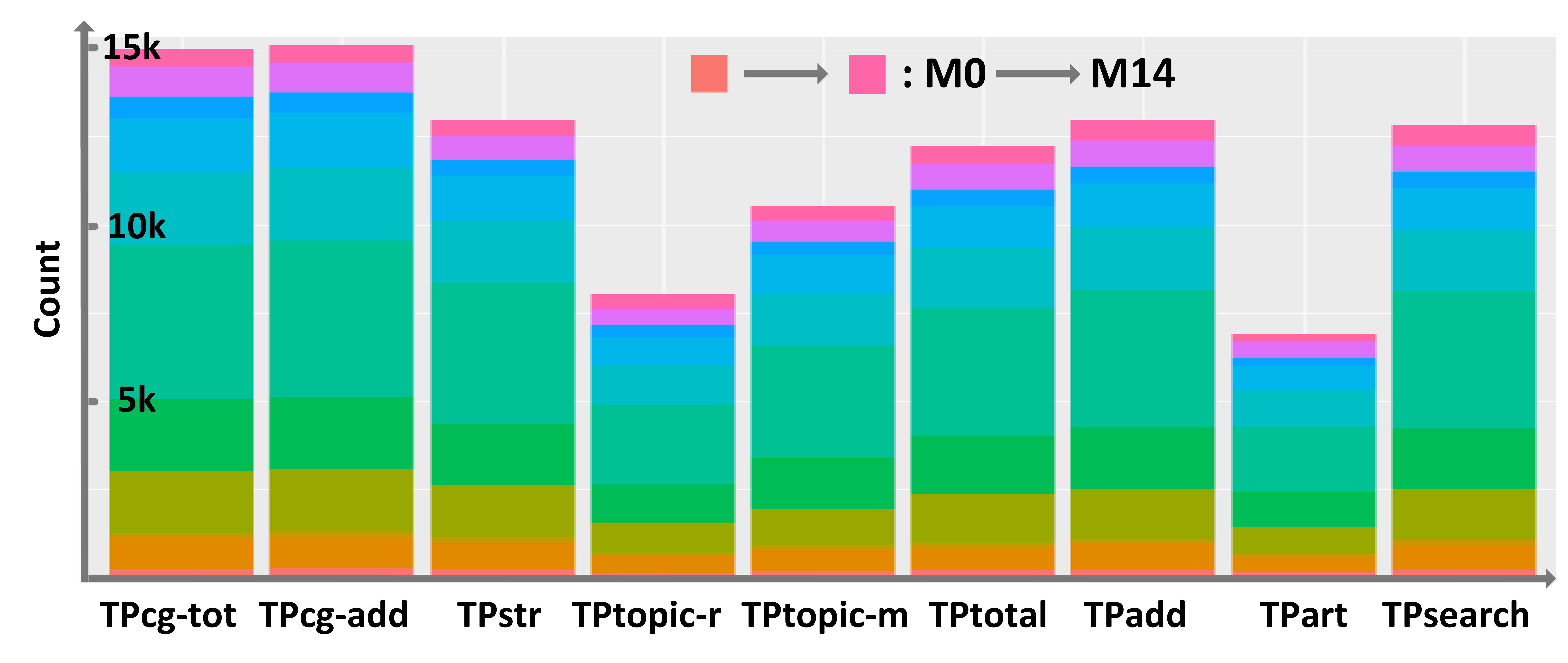}
\caption{Counts for different types of mutation faults}
\end{subfigure}
\begin{subfigure}{0.49\textwidth}
\includegraphics[width=0.98\columnwidth]{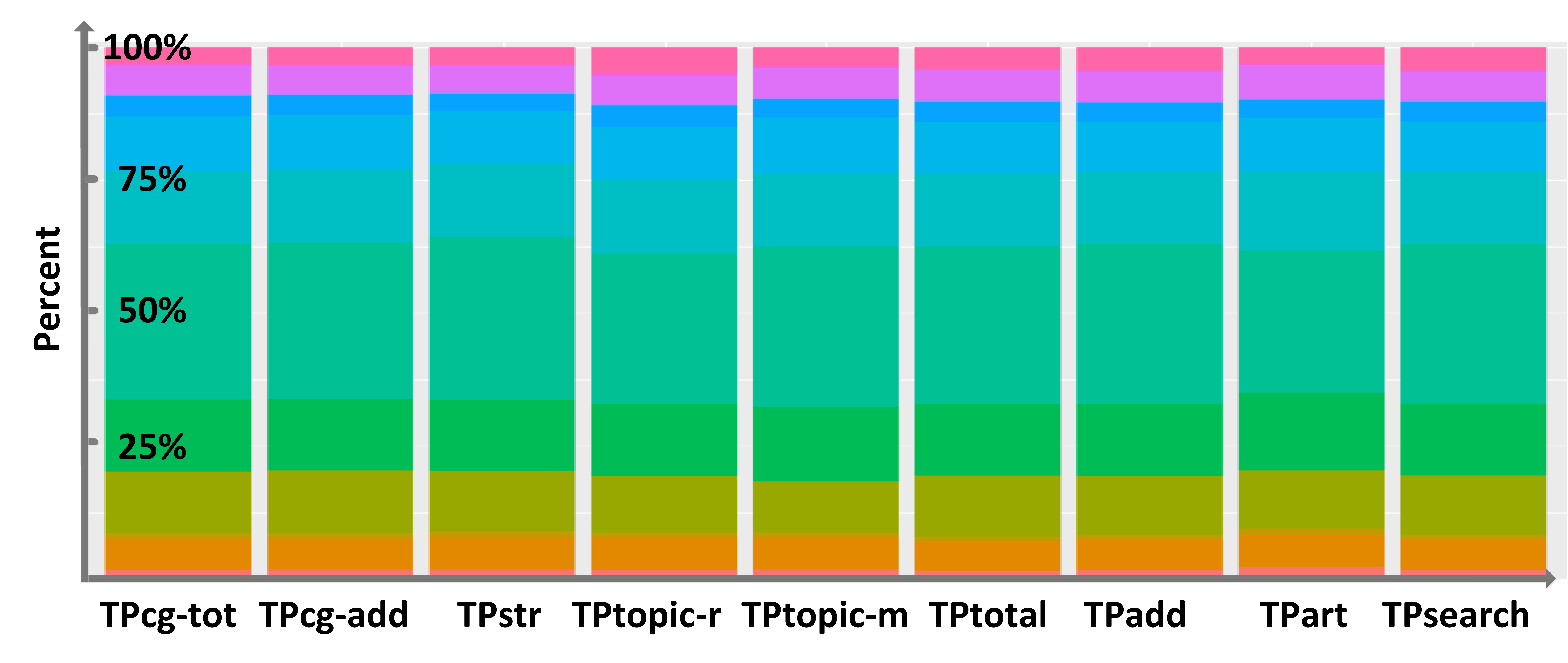}
\caption{Percentage for different types of mutation faults}
\end{subfigure}
\caption{Counts and percentage for different types of mutation faults across all subjects at cut point 10\% for class-level granularity. The types of mutation faults are classified based on the mutation operators shown in Table \ref{tab:op}.\label{fig:count-tc-10}}
\centering
\begin{subfigure}{0.49\textwidth}
\includegraphics[width=0.98\columnwidth]{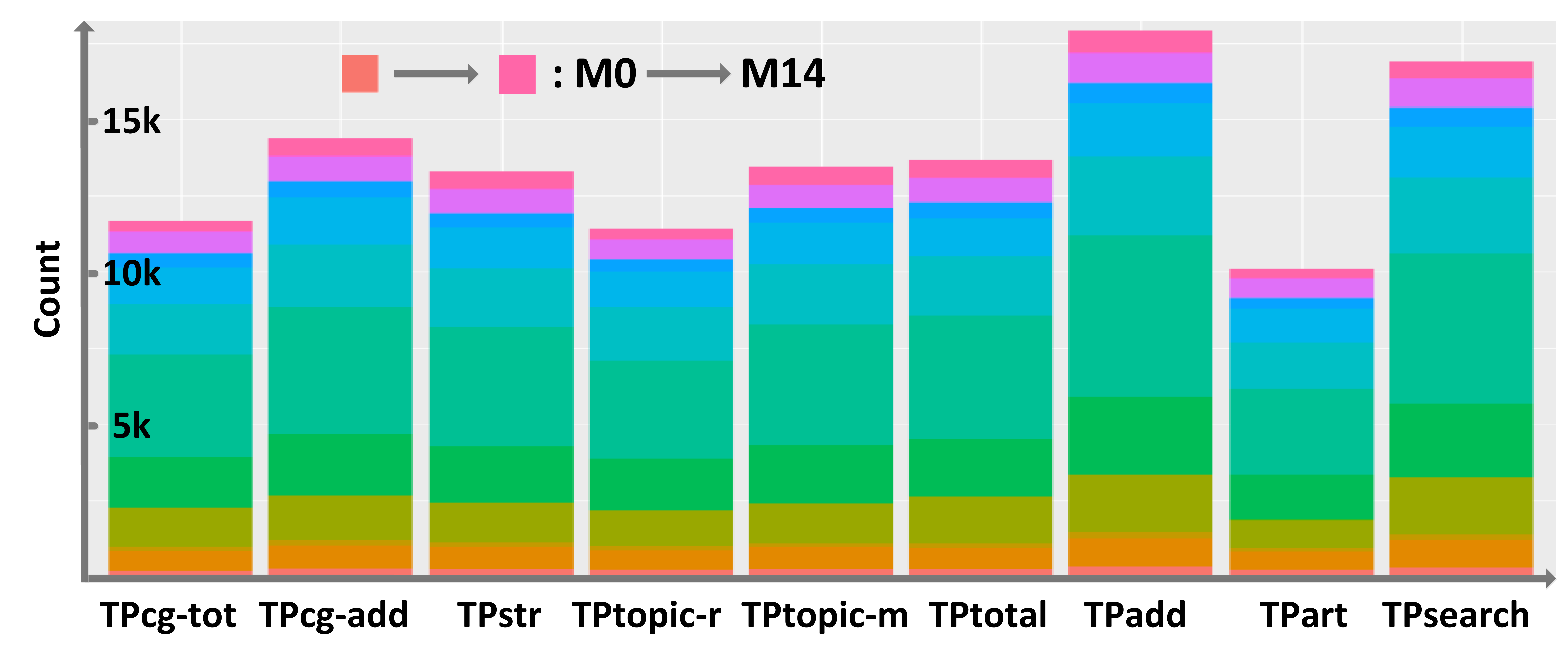}
\caption{Counts for different types of mutation faults}
\end{subfigure}
\begin{subfigure}{0.49\textwidth}
\includegraphics[width=0.98\columnwidth]{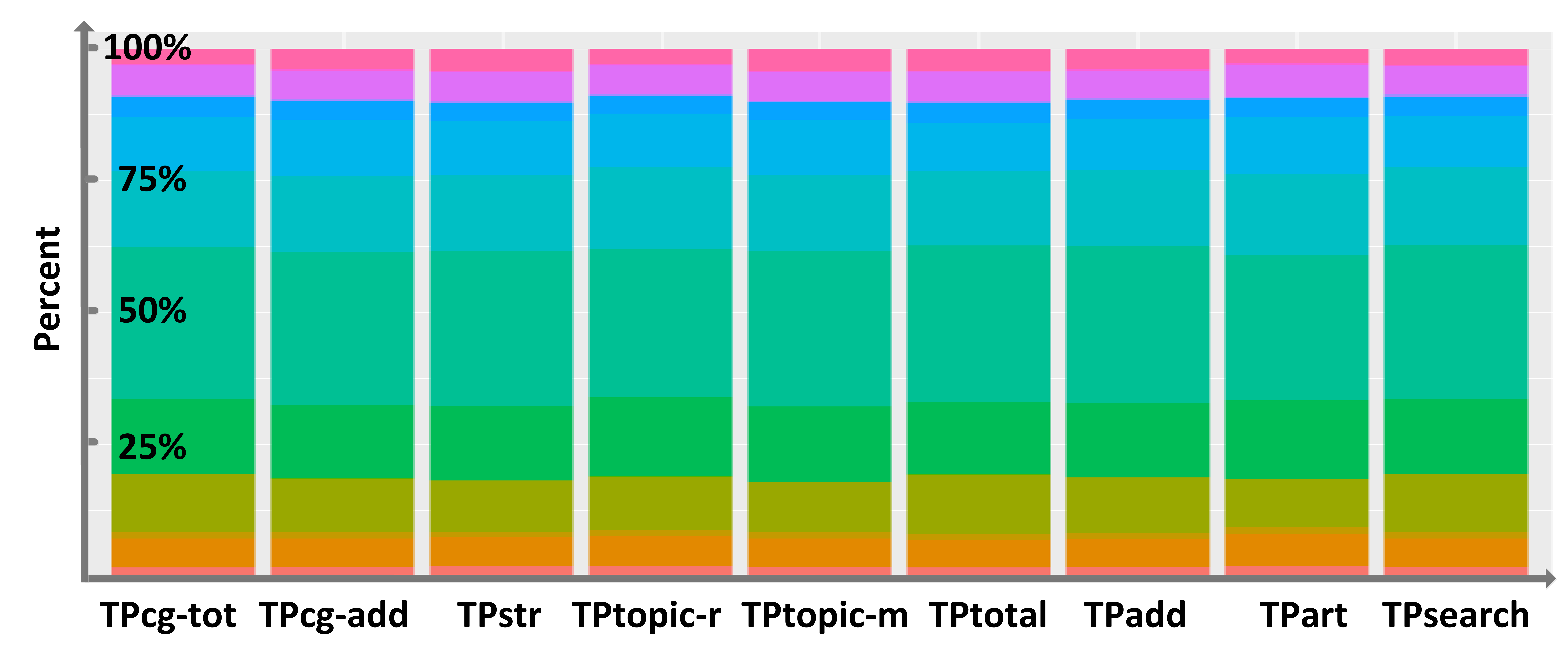}
\caption{Percentage for different types of mutation faults}
\end{subfigure}
\caption{Counts and percentage for different types of mutation faults across all subjects at cut point 10\% for method-level granularity. The types of mutation faults are classified based on the mutation operators shown in Table \ref{tab:op}.\label{fig:count-tm-10}}
\end{figure*}

\subsection{Similarity between Uncovered Faults for Different TCP techniques}

The overall results for the fault similarity analysis are shown in Figure \ref{fig:jac}. The two figures represent the results comparing the average Jaccard similarity of the studied static techniques to the studied dynamic techniques for all subject programs across 500 randomly sampled faults at different prioritization cut points. \Comment{Thus, this represents a high-level comparison of the similarity of the studied static techniques, against the studied dynamic techniques.}These results indicate that there is only a small amount of similarity between these two classifications of techniques at the higher level cut points. More specifically, for test-method level, only $\approx$ 30\% of the detected faults are similar between the two types of techniques for the top 10\% of the prioritized test cases, and at test-class level only about $\approx$ 25\% are similar for the top 10\% of prioritized test cases. This result illustrates one of the key findings of this study: The studied static and dynamic TCP techniques do not uncover similar program faults at the top cut points of prioritized test cases. The potential reason for these results is that different techniques use different types of information to prioritize test cases. For example, the studied static techniques typically aim to promote diversity between prioritized test cases using similarity/diversity metrics such as textual distance or call-graph information.  In contrast, the studied dynamic TCPs consider statement-level dynamic coverage to prioritize test cases. This finding raises interesting questions for future work regarding the possibility of combining static and dynamic information and the relative \textit{importance} of faults that differing techniques might uncover. It should be noted that different coverage granularities for dynamic TCPs may also effect the results of similarity, however we leave such an investigation for future work. From these figures we can also conclude that the techniques are slightly more similar at method level than at class level.

To further illustrate this point we calculated the Jaccard coefficients for each pair of TCPs for each subject program, and show the results in Table \ref{tab:tj-10} and Table \ref{tab:tj-50}. For each pair of techniques we group the subjects into the categories described in Section \ref{sec:study}.  Due to space limitations, we only show results for the top 10\% and top 50\% of prioritized test-cases, a complete dataset can be found at \cite{Qi:TSE17}. The results confirm the conclusions drawn from Figure \ref{fig:jac}. It is clear that when comparing the studied static and dynamic techniques, more subjects are classified into the \textit{highly-dissimilar} and \textit{dissimilar} categories at the top 10\% cut point for both of test-method and test-class levels. Another relevant conclusion that can be made is that the dissimilarity between techniques is not universal across all subjects. That is, even though two techniques may be dissimilar across several subjects, there are some cases where similarity still exists.  This suggests that only certain types of programs that exhibit different characteristics may present the opportunity of performance improvement for TCPs by using both static and dynamic information. In addition, at the cut point for the top 50\% of prioritized test cases, it is obvious that fewer subjects are classified into the \textit{highly-dissimilar} and \textit{dissimilar} categories. This is not surprising, because as the cut point increases the different techniques tend to discover more faults, limiting the potential for variance.

There are two potential reasons why we might observe higher numbers of \textit{dissimilar} faults detected at the highest cut points: 1) different types of mutants are being detected; and 2) mutants of the same type in different locations are being detected.  To investigate whether our observations are due to different fault types, we examine the counts and the percentages for different types of mutants that are detected by top 10\% test cases at both the test-class and test-method level. The results are shown in Figures \ref{fig:count-tc-10} and \ref{fig:count-tm-10}. When observing that the ratio for different types of mutation faults detected by different TCP techniques, we find that, as a whole, all TCP techniques detect a similar ratio of each mutant type, implying that mutant type is generally not the cause for the dissimilar faults at the higher cut points, but rather, mutants of the same type present different locations in source code are the more likely explanation.

\begin{framed}{
{\bf RQ$_8$: } The studied static and dynamic TCP techniques tend to discover dissimilar faults for the most highly prioritized test cases. Specifically, at the test-method level static and dynamic techniques agree only on $\approx$ 35\% of uncovered faults for the top 10\% of prioritized test cases. Additionally, a subset of subjects exhibit higher levels of detected fault similarity, suggesting that only software systems with certain characteristics may benefit from differing TCP approaches. Furthermore, the most highly prioritized test cases by different TCP techniques share similar capabilities in detecting different types of mutation faults.}
\end{framed}

\input{tables/time-sum}

\subsection{Efficiency of Static TCP Techniques}
The results of time costs for the studied static techniques at both of test-method and test-class levels are shown in
Table~\ref{tab:time-sum}. Note that, the time of pre-processing for TP$_{cg-tot}$ and TP$_{cg-add}$ are the same for both method and class levels. As the table shows, all studied techniques require similar time to pre-process the data at both method and class levels and to rank test cases on class level. But the times for prioritization are quite different at method level. We find that TP$_{cg-tot}$ and TP$_{cg-add}$ take much less time to prioritize test cases (totaling 23.78 seconds and 37.02 seconds), as compared to   TP$_{str}$ (totalling 78,835.97 seconds), TP$_{topic-r}$ (totalling 48,310.93 seconds) and TP$_{topic-m}$ (totalling 15,573.71 seconds). In particular, the following three techniques, TP$_{str}$, TP$_{topic-r}$, and TP$_{topic-m}$ take much longer time on some subjects (e.g., {\em P53} and {\em P58} ). These subjects have a large number of test cases (see Table~\ref{tab:sub}), implying that TP$_{str}$, TP$_{topic-r}$ and TP$_{topic-m}$ will take more time as the number of test cases increases. Overall, all techniques take a reasonable amount of time to preprocess data and prioritize test cases. At test-method level, TP$_{cg-tot}$ and TP$_{cg-add}$ are much more efficient. TP$_{str}$, TP$_{topic-r}$ and TP$_{topic-m}$ require more time to prioritize increasing numbers of test cases\Comment{,answering \textbf{RQ$_4$}}.

\begin{framed}{
{\bf RQ$_9$: } On test-method level, TP$_{cg-tot}$ and TP$_{cg-add}$ are the most efficient in prioritizing test cases. TP$_{str}$, TP$_{topic-r}$ and TP$_{topic-m}$ take more time when the number of test cases increases. The time of pre-processing and prioritization on test class level for all static techniques are quite similar.}
\end{framed}

%% file: tables/W-test.tex
\begin{table*}[t]
\center
\blue{
\setlength{\tabcolsep}{2pt}
\small
\caption{\small \blue{The table shows the results of Wilcoxon signed rank test on the average APFD values for each pair of TCP techniques. The techniques T1 to T9 refer to TP$_{cg-tot}$, TP$_{cg-add}$, TP$_{str}$, TP$_{topic-r}$, TP$_{topic-m}$, TP$_{total}$, TP$_{add}$, TP$_{art}$, TP$_{search}$ respectively. For each pair of TCP techniques, there are two sub-cells. The first one refers to the p-value at test-class level and the second one refers to the p-value at test-method level. The p-values are classfied into three categories, 1) p$>$0.05, 2) 0.01$<$p$<$0.05, 3) p$<$0.01. The p-values for categories  p$>$0.05 and p$<$0.01 are presented as p$>$0.05 and p$<$0.01 respectively. If a p-value is less than 0.05, the corresponding cell is shaded. } \label{tab:stat-w}}
\scalebox{0.8}{
\begin{tabular}{|c||c|c||c|c||c|c||c|c||c|c||c|c||c|c||c|c|}
\hline
&\multicolumn{2}{c||}{T2}&\multicolumn{2}{c||}{T3}&\multicolumn{2}{c||}{T4}&\multicolumn{2}{c||}{T5}&\multicolumn{2}{c||}{T6}&\multicolumn{2}{c||}{T7}&\multicolumn{2}{c||}{T8}&\multicolumn{2}{c|}{T9}\\
\hline\hline
T1&\cellcolor[gray]{0.8}0.02&\cellcolor[gray]{0.8}$<$0.01&$>$0.05&\cellcolor[gray]{0.8}$<$0.01&\cellcolor[gray]{0.8}$<$0.01&$>$0.05&\cellcolor[gray]{0.8}$<$0.01&\cellcolor[gray]{0.8}$<$0.01&\cellcolor[gray]{0.8}0.02&\cellcolor[gray]{0.8}$<$0.01&$>$0.05&\cellcolor[gray]{0.8}$<$0.01&\cellcolor[gray]{0.8}$<$0.01&$<$0.01&$>$0.05&\cellcolor[gray]{0.8}$<$0.01\\\hline
T2&\cellcolor[gray]{0.8}-&\cellcolor[gray]{0.8}-&$>$0.05&$>$0.05&\cellcolor[gray]{0.8}$<$0.01&\cellcolor[gray]{0.8}$<$0.01&\cellcolor[gray]{0.8}$<$0.01&$>$0.05&\cellcolor[gray]{0.8}$<$0.01&$>$0.05&$>$0.05&\cellcolor[gray]{0.8}$<$0.01&\cellcolor[gray]{0.8}$<$0.01&$>$0.05&$>$0.05&\cellcolor[gray]{0.8}$<$0.01\\\hline
T3&\cellcolor[gray]{0.8}-&\cellcolor[gray]{0.8}-&\cellcolor[gray]{0.8}-&\cellcolor[gray]{0.8}-&\cellcolor[gray]{0.8}$<$0.01&\cellcolor[gray]{0.8}$<$0.01&\cellcolor[gray]{0.8}$<$0.01&$>$0.05&\cellcolor[gray]{0.8}$<$0.01&$>$0.05&$>$0.05&\cellcolor[gray]{0.8}$<$0.01&\cellcolor[gray]{0.8}$<$0.01&$>$0.05&$>$0.05&\cellcolor[gray]{0.8}$<$0.01\\\hline
T4&\cellcolor[gray]{0.8}-&\cellcolor[gray]{0.8}-&\cellcolor[gray]{0.8}-&\cellcolor[gray]{0.8}-&\cellcolor[gray]{0.8}-&\cellcolor[gray]{0.8}-&\cellcolor[gray]{0.8}$<$0.01&\cellcolor[gray]{0.8}$<$0.01&\cellcolor[gray]{0.8}$<$0.01&\cellcolor[gray]{0.8}$<$0.01&\cellcolor[gray]{0.8}$<$0.01&\cellcolor[gray]{0.8}$<$0.01&0.05&\cellcolor[gray]{0.8}0.02&\cellcolor[gray]{0.8}$<$0.01&\cellcolor[gray]{0.8}$<$0.01\\\hline
T5&\cellcolor[gray]{0.8}-&\cellcolor[gray]{0.8}-&\cellcolor[gray]{0.8}-&\cellcolor[gray]{0.8}-&\cellcolor[gray]{0.8}-&\cellcolor[gray]{0.8}-&\cellcolor[gray]{0.8}-&\cellcolor[gray]{0.8}-&$>$0.05&$>$0.05&0.03&\cellcolor[gray]{0.8}$<$0.01&$<$0.01&\cellcolor[gray]{0.8}0.04&0.05&\cellcolor[gray]{0.8}$<$0.01\\\hline
T6&\cellcolor[gray]{0.8}-&\cellcolor[gray]{0.8}-&\cellcolor[gray]{0.8}-&\cellcolor[gray]{0.8}-&\cellcolor[gray]{0.8}-&\cellcolor[gray]{0.8}-&\cellcolor[gray]{0.8}-&\cellcolor[gray]{0.8}-&\cellcolor[gray]{0.8}-&\cellcolor[gray]{0.8}-&\cellcolor[gray]{0.8}$<$0.01&\cellcolor[gray]{0.8}$<$0.01&\cellcolor[gray]{0.8}$<$0.01&$>$0.05&\cellcolor[gray]{0.8}$<$0.01&\cellcolor[gray]{0.8}$<$0.01\\\hline
T7&\cellcolor[gray]{0.8}-&\cellcolor[gray]{0.8}-&\cellcolor[gray]{0.8}-&\cellcolor[gray]{0.8}-&\cellcolor[gray]{0.8}-&\cellcolor[gray]{0.8}-&\cellcolor[gray]{0.8}-&\cellcolor[gray]{0.8}-&\cellcolor[gray]{0.8}-&\cellcolor[gray]{0.8}-&\cellcolor[gray]{0.8}-&\cellcolor[gray]{0.8}-&\cellcolor[gray]{0.8}$<$0.01&\cellcolor[gray]{0.8}$<$0.01&0.04&\cellcolor[gray]{0.8}$<$0.01\\\hline
T8&\cellcolor[gray]{0.8}-&\cellcolor[gray]{0.8}-&\cellcolor[gray]{0.8}-&\cellcolor[gray]{0.8}-&\cellcolor[gray]{0.8}-&\cellcolor[gray]{0.8}-&\cellcolor[gray]{0.8}-&\cellcolor[gray]{0.8}-&\cellcolor[gray]{0.8}-&-\cellcolor[gray]{0.8}-&\cellcolor[gray]{0.8}-&\cellcolor[gray]{0.8}-&\cellcolor[gray]{0.8}-&\cellcolor[gray]{0.8}-&\cellcolor[gray]{0.8}$<$0.01&$<$0.01\\\hline
\end{tabular}
}}
\end{table*}

%% file: tables/W-test-APFDC.tex
\begin{table*}[t]
\center
\blue{
\setlength{\tabcolsep}{2pt}
\small
\caption{\small \blue{The table shows the results of Wilcoxon signed rank test on the average APFDc values for each pair of TCP techniques. 
This table follows exactly the same format as Table \ref{tab:stat-w}. \label{tab:statC-w}}}
\scalebox{0.8}{
\begin{tabular}{|c||c|c||c|c||c|c||c|c||c|c||c|c||c|c||c|c|}
\hline
&\multicolumn{2}{c||}{T2}&\multicolumn{2}{c||}{T3}&\multicolumn{2}{c||}{T4}&\multicolumn{2}{c||}{T5}&\multicolumn{2}{c||}{T6}&\multicolumn{2}{c||}{T7}&\multicolumn{2}{c||}{T8}&\multicolumn{2}{c|}{T9}\\
\hline\hline
T1&\cellcolor[gray]{0.8}0.02&\cellcolor[gray]{0.8}$<$0.01&\cellcolor[gray]{0.8}0.03&$>$0.05&\cellcolor[gray]{0.8}0.04&$>$0.05&$>$0.05&$>$0.05&$>$0.05&$>$0.05&$>$0.05&\cellcolor[gray]{0.8}$<$0.01&$>$0.05&$>$0.05&$>$0.05&\cellcolor[gray]{0.8}$<$0.01\\\hline
T2&\cellcolor[gray]{0.8}-&\cellcolor[gray]{0.8}-&$>$0.05&\cellcolor[gray]{0.8}0.01&\cellcolor[gray]{0.8}$<$0.01&\cellcolor[gray]{0.8}$<$0.01&$>$0.05&\cellcolor[gray]{0.8}0.04&$>$0.05&\cellcolor[gray]{0.8}$<$0.01&$>$0.05&$>$0.05&$>$0.05&\cellcolor[gray]{0.8}$<$0.01&$>$0.05&$>$0.05\\\hline
T3&\cellcolor[gray]{0.8}-&\cellcolor[gray]{0.8}-&\cellcolor[gray]{0.8}-&\cellcolor[gray]{0.8}-&\cellcolor[gray]{0.8}0.02&$>$0.05&$>$0.05&$>$0.05&$>$0.05&0.05&$>$0.05&\cellcolor[gray]{0.8}0.03&$>$0.05&$>$0.05&$>$0.05&\cellcolor[gray]{0.8}$<$0.01\\\hline
T4&\cellcolor[gray]{0.8}-&\cellcolor[gray]{0.8}-&\cellcolor[gray]{0.8}-&\cellcolor[gray]{0.8}-&\cellcolor[gray]{0.8}-&\cellcolor[gray]{0.8}-&\cellcolor[gray]{0.8}$<$0.01&$>$0.05&$>$0.05&$>$0.05&\cellcolor[gray]{0.8}$<$0.01&$>$0.05&$>$0.05&$>$0.05&\cellcolor[gray]{0.8}$<$0.01&\cellcolor[gray]{0.8}$<$0.01\\\hline
T5&\cellcolor[gray]{0.8}-&\cellcolor[gray]{0.8}-&\cellcolor[gray]{0.8}-&\cellcolor[gray]{0.8}-&\cellcolor[gray]{0.8}-&\cellcolor[gray]{0.8}-&\cellcolor[gray]{0.8}-&\cellcolor[gray]{0.8}-&$>$0.05&\cellcolor[gray]{0.8}0.04&$>$0.05&$>$0.05&$>$0.05&$>$0.05&$>$0.05&\cellcolor[gray]{0.8}$<$0.01\\\hline
T6&\cellcolor[gray]{0.8}-&\cellcolor[gray]{0.8}-&\cellcolor[gray]{0.8}-&\cellcolor[gray]{0.8}-&\cellcolor[gray]{0.8}-&\cellcolor[gray]{0.8}-&\cellcolor[gray]{0.8}-&\cellcolor[gray]{0.8}-&\cellcolor[gray]{0.8}-&\cellcolor[gray]{0.8}-&\cellcolor[gray]{0.8}$<$0.01&\cellcolor[gray]{0.8}$<$0.01&$>$0.05&$>$0.05&\cellcolor[gray]{0.8}0.01&\cellcolor[gray]{0.8}$<$0.01\\\hline
T7&\cellcolor[gray]{0.8}-&\cellcolor[gray]{0.8}-&\cellcolor[gray]{0.8}-&\cellcolor[gray]{0.8}-&\cellcolor[gray]{0.8}-&\cellcolor[gray]{0.8}-&\cellcolor[gray]{0.8}-&\cellcolor[gray]{0.8}-&\cellcolor[gray]{0.8}-&\cellcolor[gray]{0.8}-&\cellcolor[gray]{0.8}-&\cellcolor[gray]{0.8}-&$>$0.05&$>$0.05&$>$0.05&$>$0.05\\\hline
T8&\cellcolor[gray]{0.8}-&\cellcolor[gray]{0.8}-&\cellcolor[gray]{0.8}-&\cellcolor[gray]{0.8}-&\cellcolor[gray]{0.8}-&\cellcolor[gray]{0.8}-&\cellcolor[gray]{0.8}-&\cellcolor[gray]{0.8}-&\cellcolor[gray]{0.8}-&\cellcolor[gray]{0.8}-&\cellcolor[gray]{0.8}-&\cellcolor[gray]{0.8}-&\cellcolor[gray]{0.8}-&\cellcolor[gray]{0.8}-&$>$0.05&\cellcolor[gray]{0.8}$<$0.01\\\hline
\end{tabular}
}}
\end{table*}

%% file: tables/statisticSmallC.tex
\begin{table*}[t]
\center
\blue{
\scriptsize
\caption{\small \blue{Results for the ANOVA and Tukey HSD tests on the average APFD and APFDc values at test-class level across smaller subject programs. The last column shows the results for Kendall tau Rank Correlation Coefficient $\tau_{b}$ between the average APFDc and average APFD.\label{tab:statSmallC}}}
\begin{tabular}{|c|c|c|c|c|c|c|c|c|c|c|c|c|}
\hline
Metrics&Analysis&{TP$_{cg-tot}$}&{TP$_{cg-add}$}&{TP$_{str}$}&{TP$_{topic-r}$}&{TP$_{topic-m}$}&{TP$_{total}$}&{TP$_{add}$}&{TP$_{art}$}&{TP$_{search}$}&p-value&\taub\\
\hline\hline
\multirow{2}{*}{APFD}&Avg&0.759&0.764&0.758&0.658&0.729&0.707&0.746&0.629&0.743&\multirow{2}{*}{5.42E-8}&\multirow{4}{*}{0.5}\\\cline{2-11}
&HSD&A&A&A&BC&AB&ABC&A&C&A&&\\\cline{1-12}\cline{1-12}
\multirow{2}{*}{APFDc}&Avg&0.618&0.633&0.653&0.563&0.652&0.558&0.592&0.585&0.591&\multirow{2}{*}{0.503}&\\\cline{2-11}
&HSD&A&A&A&A&A&A&A&A&A&&\\\hline
\end{tabular}}
\end{table*}

%% file: tables/statisticLargeC.tex
\begin{table*}[t]
\center
\scriptsize
\blue{
\caption{\small \blue{Results for the ANOVA and Tukey HSD tests on the average APFD and APFDc values at test-class level across larger subject programs. The last column shows the results for Kendall tau Rank Correlation Coefficient $\tau_{b}$ between the average APFDc and average APFD.\label{tab:statLargeC}}}
\begin{tabular}{|c|c|c|c|c|c|c|c|c|c|c|c|c|}
\hline
Metrics&Analysis&{TP$_{cg-tot}$}&{TP$_{cg-add}$}&{TP$_{str}$}&{TP$_{topic-r}$}&{TP$_{topic-m}$}&{TP$_{total}$}&{TP$_{add}$}&{TP$_{art}$}&{TP$_{search}$}&p-value&\taub\\
\hline\hline
\multirow{2}{*}{APFD}&Avg&0.796&0.816&0.796&0.692&0.762&0.769&0.791&0.637&0.787&\multirow{2}{*}{7.73E-10}&\multirow{4}{*}{0.667}\\\cline{2-11}
&HSD&B&B&B&B&B&B&A&B&A&&\\\cline{1-12}\cline{1-12}
\multirow{2}{*}{APFDc}&Avg&0.686&0.724&0.681&0.586&0.661&0.670&0.708&0.640&0.706&\multirow{2}{*}{0.132}&\\\cline{2-11}
&HSD&A&A&A&A&A&A&A&A&A&&\\\hline
\end{tabular}}
\end{table*}

%% file: tables/statisticSmall.tex
\begin{table*}[t]
\center
\scriptsize
\blue{
\caption{\small \blue{Results for the ANOVA and Tukey HSD tests on the average APFD and APFDc values at test-method level across smaller subject programs. The last column shows the results for Kendall tau Rank Correlation Coefficient $\tau_{b}$ between the average APFDc and average APFD.\label{tab:statSmall}}}
\begin{tabular}{|c|c|c|c|c|c|c|c|c|c|c|c|c|}
\hline
Metrics&Analysis&{TP$_{cg-tot}$}&{TP$_{cg-add}$}&{TP$_{str}$}&{TP$_{topic-r}$}&{TP$_{topic-m}$}&{TP$_{total}$}&{TP$_{add}$}&{TP$_{art}$}&{TP$_{search}$}&p-value&\taub\\
\hline\hline
\multirow{2}{*}{APFD}&Avg&0.751&0.799&0.799&0.771&0.804&0.797&0.885&0.791&0.878&\multirow{2}{*}{2.29E-15}&\multirow{4}{*}{0.444}\\\cline{2-11}
&HSD&B&B&B&B&B&B&A&B&A&&\\\cline{1-12}\cline{1-12}
\multirow{2}{*}{APFDc}&Avg&0.604&0.700&0.670&0.655&0.673&0.605&0.657&0.645&0.696&\multirow{2}{*}{0.572}&\\\cline{2-11}
&HSD&A&A&A&A&A&A&A&A&A&&\\\hline
\end{tabular}}
\end{table*}

%% file: tables/statisticLarge.tex
\begin{table*}[t]
\center
\blue{
\scriptsize
\caption{\small \blue{Results for the ANOVA and Tukey HSD tests on the average APFD and APFDc values at test-method level across larger subject programs. The last column shows the results for Kendall tau Rank Correlation Coefficient $\tau_{b}$ between the average APFDc and average APFD.\label{tab:statLarge}}}
\begin{tabular}{|c|c|c|c|c|c|c|c|c|c|c|c|c|}
\hline
Metrics&Analysis&{TP$_{cg-tot}$}&{TP$_{cg-add}$}&{TP$_{str}$}&{TP$_{topic-r}$}&{TP$_{topic-m}$}&{TP$_{total}$}&{TP$_{add}$}&{TP$_{art}$}&{TP$_{search}$}&p-value&\taub\\
\hline\hline
\multirow{2}{*}{APFD}&Avg&0.777&0.837&0.827&0.791&0.829&0.821&0.912&0.805&0.892&\multirow{2}{*}{3.21E-12}&\multirow{4}{*}{0.333}\\\cline{2-11}
&HSD&C&BC&C&C&C&C&A&C&AB&&\\\cline{1-12}\cline{1-12}
\multirow{2}{*}{APFDc}&Avg&0.673&0.774&0.671&0.702&0.686&0.660&0.759&0.692&0.774&\multirow{2}{*}{0.086}&\\\cline{2-11}
&HSD&A&A&A&A&A&A&A&A&A&&\\\hline
\end{tabular}}
\end{table*}

%% file: evolution-Class.tex
\begin{figure*}
 \center
\begin{subfigure}{0.98\textwidth}
 \includegraphics[scale=0.5]{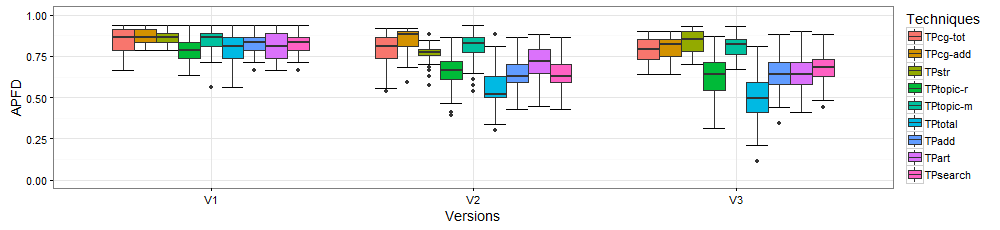}
 \caption{\blue{TCP results on geojson-jackson (APFD)}}
 \end{subfigure}
 \begin{subfigure}{0.98\textwidth}
 \includegraphics[scale=0.5]{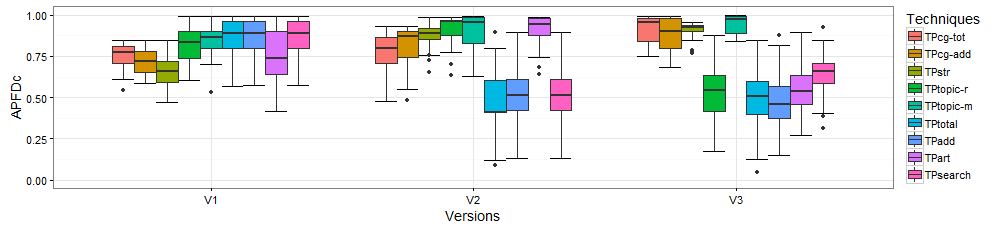}
 \caption{\blue{TCP results on geojson-jackson (APFDc)}}
 \end{subfigure}
 \begin{subfigure}{0.98\textwidth}
 \includegraphics[scale=0.5]{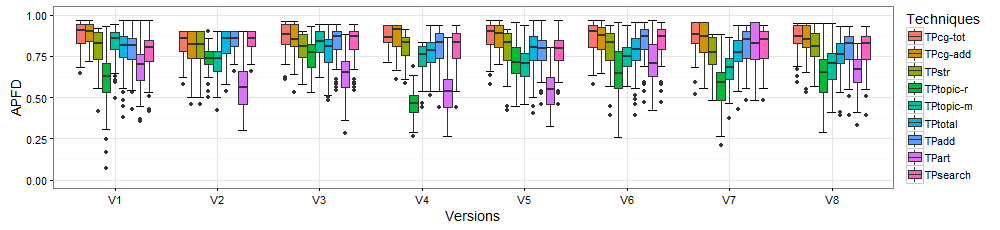}
 \caption{\blue{TCP results on javapoet (APFD)}}
 \end{subfigure}
 \begin{subfigure}{0.98\textwidth}
 \includegraphics[scale=0.5]{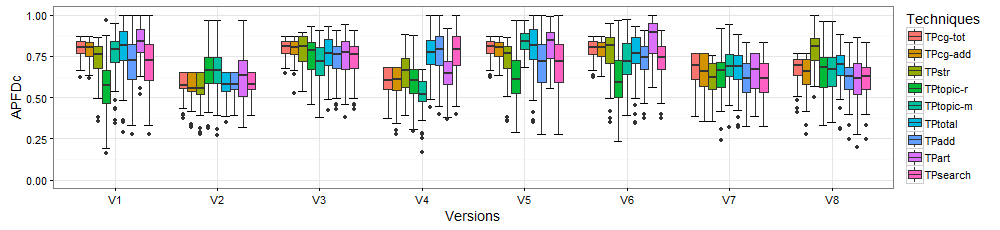}
 \caption{\blue{TCP results on javapoet (APFDc)}}
 \end{subfigure}
 \caption{\label{fig:ev-Class} \blue{Test-Class-level test prioritization in evolution}}
\end{figure*}

%% file: evolution-Method.tex
\begin{figure*}
 \center
\begin{subfigure}{0.98\textwidth}
 \includegraphics[scale=0.5]{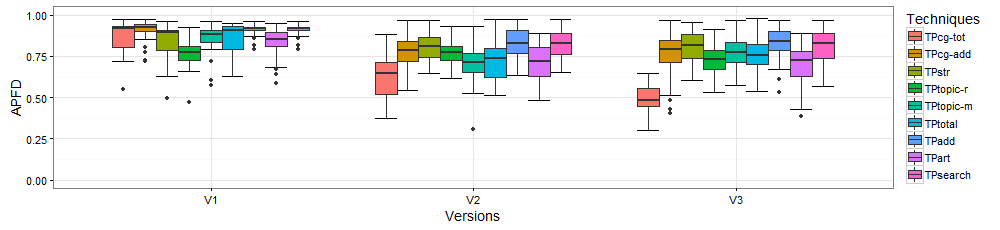}
 \caption{\blue{TCP results on geojson-jackson (APFD)}}
 \end{subfigure}
 \begin{subfigure}{0.98\textwidth}
 \includegraphics[scale=0.5]{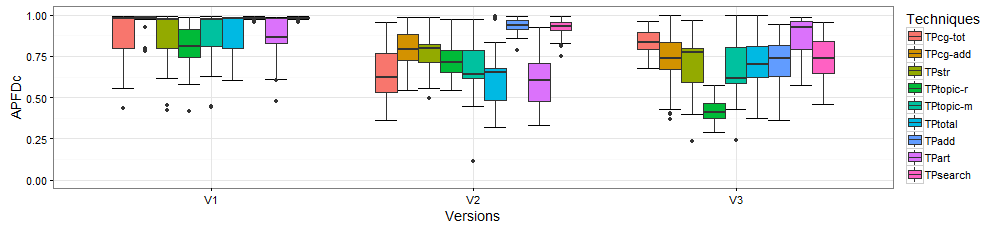}
 \caption{\blue{TCP results on geojson-jackson (APFDc)}}
 \end{subfigure}
 \begin{subfigure}{0.98\textwidth}
 \includegraphics[scale=0.5]{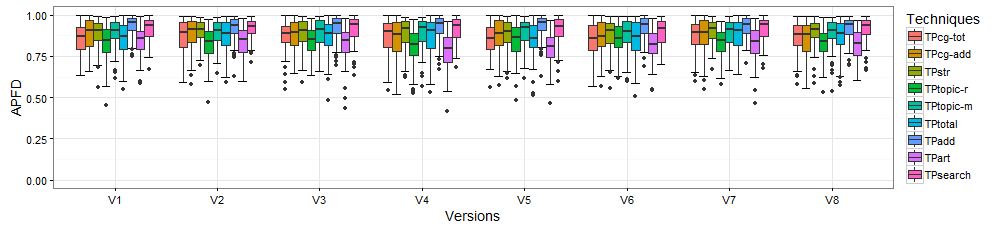}
 \caption{\blue{TCP results on javapoet (APFD)}}
 \end{subfigure}
 \begin{subfigure}{0.98\textwidth}
 \includegraphics[scale=0.5]{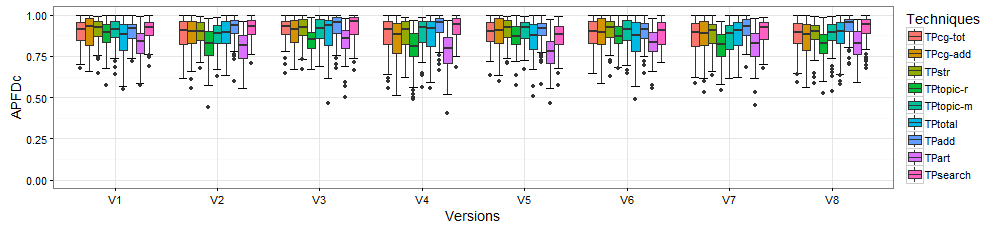}
 \caption{\blue{TCP results on javapoet (APFDc)}}
 \end{subfigure}
 \caption{\label{fig:ev-Method} \blue{Test-Method-level test prioritization in evolution}}
\end{figure*}

%% file: tables/APFD-sizes.tex
\begin{table*}[t]
\center
\scriptsize
\caption{\small Results for average APFD values on different sizes of mutation faults. The last column shows the results for Kendall tau Rank Correlation Coefficient $\tau_{b}$ between the average APFD values with different sizes of mutation faults and the average APFD values shown in Tables \ref{tab:statC} and \ref{tab:stat}.\label{tab:APFD-sizes}}
\begin{tabular}{|c|c|c|c|c|c|c|c|c|c|c|c|c|}
\hline
Granularity&Sizes&{TP$_{cg-tot}$}&{TP$_{cg-add}$}&{TP$_{str}$}&{TP$_{topic-r}$}&{TP$_{topic-m}$}&{TP$_{total}$}&{TP$_{add}$}&{TP$_{art}$}&{TP$_{search}$}&$\tau_{b}$\\
\hline\hline
\multirow{10}{*}{Test-class}
&1&0.772&0.786&0.775&0.67&0.741&0.737&0.771&0.639&0.767&0.944\\\cline{2-12}
&2&0.776&0.788&0.775&0.671&0.745&0.738&0.771&0.639&0.768&1\\\cline{2-12}
&3&0.776&0.788&0.777&0.673&0.746&0.738&0.769&0.638&0.766&0.944\\\cline{2-12}
&4&0.777&0.789&0.777&0.675&0.745&0.739&0.771&0.637&0.768&0.944\\\cline{2-12}
&5&0.777&0.789&0.778&0.675&0.746&0.739&0.771&0.635&0.767&0.944\\\cline{2-12}
&6&0.777&0.79&0.777&0.675&0.746&0.738&0.77&0.634&0.767&1\\\cline{2-12}
&7&0.778&0.79&0.777&0.675&0.746&0.738&0.769&0.633&0.766&1\\\cline{2-12}
&8&0.778&0.79&0.777&0.676&0.746&0.738&0.77&0.633&0.766&1\\\cline{2-12}
&9&0.778&0.79&0.777&0.676&0.746&0.738&0.769&0.633&0.766&1\\\cline{2-12}
&10&0.778&0.79&0.777&0.676&0.747&0.738&0.77&0.634&0.766&1\\\cline{2-12}\hline\hline
\multirow{10}{*}{Test-method}
&1&0.759&0.82&0.813&0.777&0.814&0.807&0.901&0.798&0.885&1\\\cline{2-12}
&2&0.763&0.82&0.816&0.781&0.818&0.809&0.902&0.802&0.887&1\\\cline{2-12}
&3&0.763&0.818&0.814&0.78&0.815&0.81&0.9&0.8&0.886&1\\\cline{2-12}
&4&0.764&0.819&0.814&0.782&0.817&0.811&0.901&0.802&0.887&1\\\cline{2-12}
&5&0.764&0.818&0.814&0.782&0.818&0.811&0.9&0.8&0.887&1\\\cline{2-12}
&6&0.762&0.817&0.813&0.781&0.816&0.809&0.9&0.8&0.886&1\\\cline{2-12}
&7&0.763&0.817&0.813&0.781&0.816&0.81&0.9&0.8&0.886&1\\\cline{2-12}
&8&0.763&0.818&0.812&0.781&0.816&0.81&0.899&0.8&0.886&1\\\cline{2-12}
&9&0.763&0.818&0.812&0.781&0.816&0.81&0.899&0.799&0.885&1\\\cline{2-12}
&10&0.764&0.818&0.813&0.781&0.816&0.809&0.899&0.799&0.885&1\\\cline{2-12}\hline
\end{tabular}
\end{table*} 

%% file: tables/APFDC-sizes.tex
\begin{table*}[t]
\center
\scriptsize
\caption{\small \blue{Results for average APFDc values on different sizes of mutation faults. This table follows the same format as Table \ref{tab:APFD-sizes}.}\label{tab:APFDC-sizes}}
\begin{tabular}{|c|c|c|c|c|c|c|c|c|c|c|c|c|}
\hline
Granularity&Sizes&{TP$_{cg-tot}$}&{TP$_{cg-add}$}&{TP$_{str}$}&{TP$_{topic-r}$}&{TP$_{topic-m}$}&{TP$_{total}$}&{TP$_{add}$}&{TP$_{art}$}&{TP$_{search}$}&$\tau_{b}$\\
\hline\hline
\multirow{10}{*}{Test-class}
&1&0.655&0.681&0.668&0.572&0.658&0.619&0.653&0.622&0.652&0.944\\\cline{2-12}
&2&0.653&0.679&0.668&0.571&0.66&0.616&0.652&0.621&0.652&0.944\\\cline{2-12}
&3&0.653&0.679&0.669&0.574&0.66&0.617&0.651&0.619&0.65&0.944\\\cline{2-12}
&4&0.652&0.678&0.667&0.575&0.657&0.617&0.652&0.615&0.651&1\\\cline{2-12}
&5&0.652&0.678&0.668&0.575&0.659&0.616&0.651&0.614&0.651&1\\\cline{2-12}
&6&0.653&0.68&0.667&0.575&0.658&0.616&0.65&0.613&0.65&1\\\cline{2-12}
&7&0.653&0.68&0.667&0.575&0.658&0.616&0.65&0.613&0.65&1\\\cline{2-12}
&8&0.654&0.681&0.667&0.576&0.659&0.617&0.651&0.613&0.65&1\\\cline{2-12}
&9&0.654&0.68&0.668&0.576&0.659&0.617&0.651&0.613&0.65&1\\\cline{2-12}
&10&0.654&0.68&0.668&0.576&0.659&0.616&0.651&0.613&0.65&1\\\cline{2-12}\hline\hline
\multirow{10}{*}{Test-method}
&1&0.637&0.745&0.671&0.681&0.681&0.634&0.715&0.675&0.739&0.889\\\cline{2-12}
&2&0.639&0.739&0.675&0.683&0.684&0.634&0.714&0.675&0.739&0.944\\\cline{2-12}
&3&0.639&0.738&0.672&0.681&0.679&0.634&0.711&0.672&0.738&0.889\\\cline{2-12}
&4&0.638&0.737&0.671&0.68&0.68&0.634&0.711&0.672&0.737&0.944\\\cline{2-12}
&5&0.638&0.737&0.672&0.68&0.681&0.635&0.711&0.671&0.738&0.944\\\cline{2-12}
&6&0.636&0.738&0.671&0.678&0.68&0.634&0.711&0.671&0.737&0.944\\\cline{2-12}
&7&0.637&0.738&0.671&0.679&0.68&0.635&0.711&0.671&0.737&1\\\cline{2-12}
&8&0.638&0.739&0.67&0.679&0.68&0.635&0.71&0.671&0.737&0.944\\\cline{2-12}
&9&0.638&0.738&0.67&0.679&0.679&0.635&0.71&0.67&0.737&0.944\\\cline{2-12}
&10&0.638&0.737&0.67&0.678&0.679&0.634&0.709&0.67&0.736&1\\\cline{2-12}\hline
\end{tabular}
\end{table*} 

%% file: tables/APFD-types.tex
\begin{table*}[t]
\center
\scriptsize
\caption{\small Results for average APFD values on different types of mutation faults. The last column shows the results for Kendall tau Rank Correlation Coefficient $\tau_{b}$ between the average APFD values with different types of mutation faults and the average APFD values shown in Tables \ref{tab:statC} and \ref{tab:stat}.\label{tab:APFD-types}}
\begin{tabular}{|c|c|c|c|c|c|c|c|c|c|c|c|c|}
\hline
Gra.&Types&{TP$_{cg-tot}$}&{TP$_{cg-add}$}&{TP$_{str}$}&{TP$_{topic-r}$}&{TP$_{topic-m}$}&{TP$_{total}$}&{TP$_{add}$}&{TP$_{art}$}&{TP$_{search}$}&$\tau_{b}$\\
\hline\hline
\multirow{15}{*}{Class}
&NegateConditionals&0.785&0.796&0.792&0.699&0.762&0.749&0.788&0.677&0.784&0.889\\\cline{2-12}
&RemoveConditional&0.792&0.803&0.794&0.703&0.764&0.755&0.792&0.679&0.788&0.944\\\cline{2-12}
&ConstructorCall&0.784&0.797&0.787&0.682&0.757&0.74&0.772&0.663&0.766&0.944\\\cline{2-12}
&NonVoidMethodCall&0.774&0.783&0.774&0.668&0.74&0.731&0.761&0.626&0.756&1\\\cline{2-12}
&Math&0.782&0.787&0.776&0.685&0.748&0.706&0.775&0.657&0.772&1\\\cline{2-12}
&MemberVariable&0.798&0.816&0.772&0.69&0.752&0.776&0.794&0.676&0.79&0.778\\\cline{2-12}
&InlineConstant&0.76&0.778&0.776&0.687&0.752&0.707&0.752&0.641&0.75&0.889\\\cline{2-12}
&Increments&0.791&0.815&0.792&0.725&0.784&0.726&0.8&0.7&0.796&0.722\\\cline{2-12}
&ArgumentPropagation&0.775&0.787&0.775&0.676&0.751&0.738&0.769&0.624&0.766&1\\\cline{2-12}
&ConditionalsBoundary&0.787&0.81&0.809&0.709&0.776&0.737&0.78&0.69&0.778&0.944\\\cline{2-12}
&Switch&0.859&0.838&0.882&0.822&0.856&0.849&0.867&0.757&0.864&0.5\\\cline{2-12}
&VoidMethodCall&0.782&0.781&0.771&0.659&0.72&0.757&0.749&0.623&0.748&0.778\\\cline{2-12}
&InvertNegs&0.744&0.805&0.849&0.669&0.738&0.63&0.757&0.726&0.743&0.667\\\cline{2-12}
&ReturnVals&0.781&0.802&0.779&0.671&0.748&0.732&0.762&0.651&0.759&1\\\cline{2-12}
&RemoveIncrements&0.755&0.797&0.761&0.684&0.738&0.685&0.762&0.645&0.759&0.778\\\cline{2-12}
\hline\hline
\multirow{15}{*}{Method}
&NegateConditionals&0.787&0.842&0.845&0.809&0.851&0.829&0.921&0.828&0.911&0.889\\\cline{2-12}
&RemoveConditional&0.792&0.846&0.848&0.813&0.852&0.834&0.925&0.832&0.914&0.889\\\cline{2-12}
&ConstructorCall&0.756&0.816&0.8&0.782&0.806&0.791&0.886&0.808&0.873&0.833\\\cline{2-12}
&NonVoidMethodCall&0.755&0.809&0.811&0.766&0.815&0.804&0.891&0.783&0.874&0.889\\\cline{2-12}
&Math&0.745&0.801&0.789&0.776&0.795&0.777&0.902&0.809&0.882&0.778\\\cline{2-12}
&MemberVariable&0.799&0.858&0.837&0.809&0.843&0.838&0.914&0.832&0.905&0.944\\\cline{2-12}
&InlineConstant&0.735&0.777&0.785&0.766&0.788&0.774&0.881&0.794&0.869&0.667\\\cline{2-12}
&Increments&0.771&0.853&0.857&0.835&0.865&0.81&0.942&0.854&0.926&0.722\\\cline{2-12}
&ArgumentPropagation&0.752&0.821&0.824&0.768&0.829&0.81&0.899&0.793&0.881&0.889\\\cline{2-12}
&ConditionalsBoundary&0.761&0.825&0.827&0.806&0.833&0.809&0.901&0.819&0.89&0.833\\\cline{2-12}
&Switch&0.86&0.881&0.932&0.871&0.946&0.902&0.968&0.892&0.952&0.778\\\cline{2-12}
&VoidMethodCall&0.771&0.805&0.792&0.766&0.8&0.81&0.871&0.779&0.862&0.778\\\cline{2-12}
&InvertNegs&0.752&0.843&0.813&0.69&0.784&0.726&0.849&0.799&0.812&0.611\\\cline{2-12}
&ReturnVals&0.766&0.842&0.82&0.785&0.819&0.792&0.889&0.803&0.876&0.889\\\cline{2-12}
&RemoveIncrements&0.737&0.84&0.861&0.807&0.868&0.777&0.935&0.846&0.907&0.722\\\cline{2-12}
\hline
\end{tabular}
\end{table*} 

%% file: tables/APFDC-types.tex
\begin{table*}[t]
\center
\scriptsize
\caption{\small Results for average APFDc values on different types of mutation faults. The last column shows the results for Kendall tau Rank Correlation Coefficient $\tau_{b}$ between the average APFDc values with different types of mutation faults and the average APFDc values shown in Tables \ref{tab:statC} and \ref{tab:stat}.\label{tab:APFDC-types}}
\begin{tabular}{|c|c|c|c|c|c|c|c|c|c|c|c|c|}
\hline
Gra.&Types&{TP$_{cg-tot}$}&{TP$_{cg-add}$}&{TP$_{str}$}&{TP$_{topic-r}$}&{TP$_{topic-m}$}&{TP$_{total}$}&{TP$_{add}$}&{TP$_{art}$}&{TP$_{search}$}&$\tau_{b}$\\
\hline\hline
\multirow{15}{*}{Class}
&NegateConditionals&0.646&0.67&0.671&0.591&0.666&0.614&0.653&0.639&0.652&0.778\\\cline{2-12}
&RemoveConditional&0.655&0.677&0.673&0.594&0.668&0.62&0.659&0.64&0.658&0.833\\\cline{2-12}
&ConstructorCall&0.651&0.682&0.672&0.575&0.664&0.609&0.644&0.627&0.64&0.944\\\cline{2-12}
&NonVoidMethodCall&0.642&0.665&0.652&0.556&0.645&0.598&0.628&0.593&0.625&1\\\cline{2-12}
&Math&0.658&0.66&0.653&0.576&0.636&0.559&0.635&0.611&0.634&0.778\\\cline{2-12}
&MemberVariable&0.668&0.699&0.656&0.587&0.66&0.649&0.671&0.641&0.667&0.556\\\cline{2-12}
&InlineConstant&0.62&0.647&0.651&0.57&0.644&0.567&0.617&0.607&0.619&0.778\\\cline{2-12}
&Increments&0.636&0.663&0.655&0.584&0.657&0.584&0.639&0.633&0.639&0.667\\\cline{2-12}
&ArgumentPropagation&0.655&0.668&0.653&0.571&0.66&0.613&0.644&0.593&0.644&0.889\\\cline{2-12}
&ConditionalsBoundary&0.64&0.673&0.674&0.591&0.655&0.604&0.643&0.639&0.647&0.722\\\cline{2-12}
&Switch&0.806&0.775&0.803&0.663&0.779&0.736&0.763&0.822&0.759&0.333\\\cline{2-12}
&VoidMethodCall&0.628&0.631&0.624&0.519&0.602&0.606&0.602&0.579&0.605&0.611\\\cline{2-12}
&InvertNegs&0.537&0.579&0.705&0.498&0.666&0.472&0.746&0.747&0.714&0\\\cline{2-12}
&ReturnVals&0.652&0.685&0.661&0.56&0.649&0.603&0.636&0.619&0.635&0.889\\\cline{2-12}
&RemoveIncrements&0.605&0.639&0.664&0.546&0.628&0.544&0.602&0.649&0.602&0.5\\\cline{2-12}
\hline\hline
\multirow{15}{*}{Method}
&NegateConditionals&0.647&0.749&0.691&0.699&0.7&0.637&0.724&0.694&0.757&0.889\\\cline{2-12}
&RemoveConditional&0.653&0.753&0.696&0.704&0.706&0.643&0.73&0.701&0.762&0.889\\\cline{2-12}
&ConstructorCall&0.632&0.728&0.651&0.667&0.658&0.621&0.698&0.68&0.715&0.778\\\cline{2-12}
&NonVoidMethodCall&0.617&0.716&0.653&0.648&0.664&0.619&0.688&0.637&0.709&0.889\\\cline{2-12}
&Math&0.631&0.723&0.655&0.679&0.683&0.581&0.722&0.709&0.748&0.778\\\cline{2-12}
&MemberVariable&0.673&0.773&0.701&0.708&0.703&0.657&0.731&0.715&0.761&0.778\\\cline{2-12}
&InlineConstant&0.598&0.683&0.622&0.654&0.636&0.573&0.67&0.659&0.699&0.722\\\cline{2-12}
&Increments&0.606&0.739&0.671&0.706&0.695&0.611&0.731&0.713&0.763&0.667\\\cline{2-12}
&ArgumentPropagation&0.609&0.709&0.662&0.646&0.674&0.611&0.698&0.64&0.713&0.833\\\cline{2-12}
&ConditionalsBoundary&0.605&0.716&0.641&0.682&0.661&0.612&0.695&0.68&0.73&0.722\\\cline{2-12}
&Switch&0.752&0.787&0.786&0.78&0.796&0.768&0.812&0.777&0.839&0.722\\\cline{2-12}
&VoidMethodCall&0.626&0.683&0.615&0.619&0.63&0.599&0.635&0.613&0.666&0.833\\\cline{2-12}
&InvertNegs&0.7&0.816&0.6&0.537&0.655&0.651&0.66&0.713&0.752&0.389\\\cline{2-12}
&ReturnVals&0.63&0.742&0.661&0.671&0.669&0.608&0.687&0.664&0.711&0.889\\\cline{2-12}
&RemoveIncrements&0.597&0.752&0.699&0.697&0.726&0.575&0.709&0.73&0.743&0.667\\\cline{2-12}
\hline
\end{tabular}
\end{table*} 

%% file: tables/JD-10percent.tex
\begin{table*}[t]
\center
\tiny
\setlength{\tabcolsep}{4.8pt}
\caption{\small \label{tab:tj-10}The tables show the classification of subjects on different granularities using Jaccard distance. The four values in each cell are the numbers of subject projects, the faults of which detected by two techniques are highly dissimilar, dissimilar, similar and highly similar respectively. The technique enumeration is consistent with Table \ref{tab:stat-w}. }
\begin{subtable}{0.99\textwidth}
  \caption{This table shows the classification of subjects at the cut point 10\% on test-class level. \label{tab:tcj-10}}
\begin{tabular}{|c|p{1.2mm}p{1.2mm}p{1.2mm}p{1.2mm}|p{1.2mm}p{1.2mm}p{1.2mm}p{1.2mm}|p{1.2mm}p{1.2mm}p{1.2mm}p{1.2mm}|p{1.2mm}p{1.2mm}p{1.2mm}p{1.2mm}|p{1.2mm}p{1.2mm}p{1.2mm}p{1.2mm}|p{1.2mm}p{1.2mm}p{1.2mm}p{1.2mm}|p{1.2mm}p{1.2mm}p{1.2mm}p{1.2mm}|p{1.2mm}p{1.2mm}p{1.2mm}p{1.2mm}|p{1.2mm}p{1.2mm}p{1.2mm}p{1.2mm}|}
\hline
&\multicolumn{4}{c|}{T1}&\multicolumn{4}{c|}{T2}&\multicolumn{4}{c|}{T3}&\multicolumn{4}{c|}{T4}&\multicolumn{4}{c|}{T5}&\multicolumn{4}{c|}{T6}&\multicolumn{4}{c|}{T7}&\multicolumn{4}{c|}{T8}&\multicolumn{4}{c|}{T9}\\\hline
TP1&--&--&--&--&3&2&16&37&11&13&18&16&28&20&3&7&14&18&17&9&11&19&13&15&15&14&16&13&32&18&5&3&15&15&13&15\\\hline
TP2&3&2&16&37&--&--&--&--&11&14&15&18&27&23&3&5&13&24&14&7&11&20&14&13&14&16&13&15&30&23&4&1&13&17&10&18\\\hline
TP3&11&13&18&16&11&14&15&18&--&--&--&--&30&13&10&5&12&12&14&20&20&15&12&11&18&18&13&9&33&15&6&4&18&18&12&10\\\hline
TP4&28&20&3&7&27&23&3&5&30&13&10&5&--&--&--&--&14&15&12&17&26&17&11&4&24&15&15&4&31&13&7&7&24&16&13&5\\\hline
TP5&14&18&17&9&13&24&14&7&12&12&14&20&14&15&12&17&--&--&--&--&19&24&7&8&21&20&10&7&30&16&7&5&21&20&9&8\\\hline
TP6&11&19&13&15&11&20&14&13&20&15&12&11&26&17&11&4&19&24&7&8&--&--&--&--&2&13&11&32&28&13&9&8&2&12&14&30\\\hline
TP7&15&14&16&13&14&16&13&15&18&18&13&9&24&15&15&4&21&20&10&7&2&13&11&32&--&--&--&--&25&16&13&4&0&0&2&56\\\hline
TP8&32&18&5&3&30&23&4&1&33&15&6&4&31&13&7&7&30&16&7&5&28&13&9&8&25&16&13&4&--&--&--&--&25&15&15&3\\\hline
TP9&15&15&13&15&13&17&10&18&18&18&12&10&24&16&13&5&21&20&9&8&2&12&14&30&0&0&2&56&25&15&15&3&--&--&--&--\\\hline
\hline
Total&129&119&101&115&122&139&89&114&153&118&100&93&204&132&74&54&144&149&90&81&119&133&91&121&119&112&93&140&234&129&66&35&118&113&88&145\\\hline
\end{tabular}
\end{subtable}
\begin{subtable}{0.99\textwidth}
 \caption{This table shows the classification of subjects at the cut point 10\% on test-method level. \label{tab:tmj-10}}
\begin{tabular}{|c|p{1.2mm}p{1.2mm}p{1.2mm}p{1.2mm}|p{1.2mm}p{1.2mm}p{1.2mm}p{1.2mm}|p{1.2mm}p{1.2mm}p{1.2mm}p{1.2mm}|p{1.2mm}p{1.2mm}p{1.2mm}p{1.2mm}|p{1.2mm}p{1.2mm}p{1.2mm}p{1.2mm}|p{1.2mm}p{1.2mm}p{1.2mm}p{1.2mm}|p{1.2mm}p{1.2mm}p{1.2mm}p{1.2mm}|p{1.2mm}p{1.2mm}p{1.2mm}p{1.2mm}|p{1.2mm}p{1.2mm}p{1.2mm}p{1.2mm}|}
\hline
&\multicolumn{4}{c|}{T1}&\multicolumn{4}{c|}{T2}&\multicolumn{4}{c|}{T3}&\multicolumn{4}{c|}{T4}&\multicolumn{4}{c|}{T5}&\multicolumn{4}{c|}{T6}&\multicolumn{4}{c|}{T7}&\multicolumn{4}{c|}{T8}&\multicolumn{4}{c|}{T9}\\\hline
TP1&--&--&--&--&3&14&28&13&11&23&17&7&11&29&15&3&12&20&20&6&6&14&19&19&6&19&22&11&21&23&11&3&3&21&20&14\\\hline
TP2&3&14&28&13&--&--&--&--&7&18&24&9&5&22&26&5&6&19&28&5&1&21&25&11&2&14&23&19&14&21&19&4&3&14&25&16\\\hline
TP3&11&23&17&7&7&18&24&9&--&--&--&--&4&21&27&6&0&3&17&38&7&16&23&12&4&12&26&16&15&22&20&1&5&12&29&12\\\hline
TP4&11&29&15&3&5&22&26&5&4&21&27&6&--&--&--&--&6&22&26&4&7&27&21&3&5&26&23&4&12&25&19&2&4&26&22&6\\\hline
TP5&12&20&20&6&6&19&28&5&0&3&17&38&6&22&26&4&--&--&--&--&7&17&24&10&6&5&34&13&13&24&20&1&7&7&34&10\\\hline
TP6&6&14&19&19&1&21&25&11&7&16&23&12&7&27&21&3&7&17&24&10&--&--&--&--&1&11&29&17&19&21&16&2&2&11&26&19\\\hline
TP7&6&19&22&11&2&14&23&19&4&12&26&16&5&26&23&4&6&5&34&13&1&11&29&17&--&--&--&--&10&19&26&3&1&3&6&48\\\hline
TP8&21&23&11&3&14&21&19&4&15&22&20&1&12&25&19&2&13&24&20&1&19&21&16&2&10&19&26&3&--&--&--&--&13&17&23&5\\\hline
TP9&3&21&20&14&3&14&25&16&5&12&29&12&4&26&22&6&7&7&34&10&2&11&26&19&1&3&6&48&13&17&23&5&--&--&--&--\\\hline
\hline
Total&73&163&152&76&41&143&198&82&53&127&183&101&54&198&179&33&57&117&203&87&50&138&183&93&35&109&189&131&117&172&154&21&38&111&185&130\\\hline
\end{tabular}
\end{subtable}
\end{table*}

%% file: tables/JD-50percent.tex
\begin{table*}[t]
\center
\tiny
\setlength{\tabcolsep}{4.8pt}
\caption{\small\label{tab:tj-50} The tables show the classification of subjects on different granularities using Jaccard distance. The four values in each cell are the numbers of subject projects, the faults of which detected by two techniques are highly dissimilar, dissimilar, similar and highly similar respectively. The technique enumeration is consistent with Table \ref{tab:stat-w}. }
\begin{subtable}{0.99\textwidth}
  \caption{This table shows the classification of subjects at the cut point 50\% on test-class level. \label{tab:tcj-50}}
\begin{tabular}{|c|p{1.2mm}p{1.2mm}p{1.2mm}p{1.2mm}|p{1.2mm}p{1.2mm}p{1.2mm}p{1.2mm}|p{1.2mm}p{1.2mm}p{1.2mm}p{1.2mm}|p{1.2mm}p{1.2mm}p{1.2mm}p{1.2mm}|p{1.2mm}p{1.2mm}p{1.2mm}p{1.2mm}|p{1.2mm}p{1.2mm}p{1.2mm}p{1.2mm}|p{1.2mm}p{1.2mm}p{1.2mm}p{1.2mm}|p{1.2mm}p{1.2mm}p{1.2mm}p{1.2mm}|p{1.2mm}p{1.2mm}p{1.2mm}p{1.2mm}|}
\hline
&\multicolumn{4}{c|}{T1}&\multicolumn{4}{c|}{T2}&\multicolumn{4}{c|}{T3}&\multicolumn{4}{c|}{T4}&\multicolumn{4}{c|}{T5}&\multicolumn{4}{c|}{T6}&\multicolumn{4}{c|}{T7}&\multicolumn{4}{c|}{T8}&\multicolumn{4}{c|}{T9}\\\hline
TP1&--&--&--&--&0&2&10&46&0&1&10&47&1&6&30&21&1&3&17&37&0&6&17&35&0&2&22&34&2&9&29&18&0&2&22&34\\\hline
TP2&0&2&10&46&--&--&--&--&1&0&9&48&0&6&28&24&0&2&17&39&0&2&23&33&0&1&17&40&2&7&24&25&0&1&16&41\\\hline
TP3&0&1&10&47&1&0&9&48&--&--&--&--&1&6&23&28&0&2&11&45&1&4&19&34&1&1&16&40&2&9&21&26&1&1&16&40\\\hline
TP4&1&6&30&21&0&6&28&24&1&6&23&28&--&--&--&--&0&7&20&31&0&7&32&19&1&3&29&25&4&9&19&26&1&3&29&25\\\hline
TP5&1&3&17&37&0&2&17&39&0&2&11&45&0&7&20&31&--&--&--&--&0&7&20&31&0&4&17&37&3&7&25&23&0&4&18&36\\\hline
TP6&0&6&17&35&0&2&23&33&1&4&19&34&0&7&32&19&0&7&20&31&--&--&--&--&0&2&18&38&2&15&28&13&0&2&18&38\\\hline
TP7&0&2&22&34&0&1&17&40&1&1&16&40&1&3&29&25&0&4&17&37&0&2&18&38&--&--&--&--&2&11&13&32&0&0&0&58\\\hline
TP8&2&9&29&18&2&7&24&25&2&9&21&26&4&9&19&26&3&7&25&23&2&15&28&13&2&11&13&32&--&--&--&--&2&10&13&33\\\hline
TP9&0&2&22&34&0&1&16&41&1&1&16&40&1&3&29&25&0&4&18&36&0&2&18&38&0&0&0&58&2&10&13&33&--&--&--&--\\\hline
\hline
Total&4&31&157&272&3&21&144&296&7&24&125&308&8&47&210&199&4&36&145&279&3&45&175&241&4&24&132&304&19&77&172&196&4&23&132&305\\\hline
\end{tabular}
\end{subtable}
\begin{subtable}{0.99\textwidth}
\caption{This table shows the classification of subjects at the cut point 50\% on test-method level. \label{tab:tmj-50}}
\begin{tabular}{|c|p{1.2mm}p{1.2mm}p{1.2mm}p{1.2mm}|p{1.2mm}p{1.2mm}p{1.2mm}p{1.2mm}|p{1.2mm}p{1.2mm}p{1.2mm}p{1.2mm}|p{1.2mm}p{1.2mm}p{1.2mm}p{1.2mm}|p{1.2mm}p{1.2mm}p{1.2mm}p{1.2mm}|p{1.2mm}p{1.2mm}p{1.2mm}p{1.2mm}|p{1.2mm}p{1.2mm}p{1.2mm}p{1.2mm}|p{1.2mm}p{1.2mm}p{1.2mm}p{1.2mm}|p{1.2mm}p{1.2mm}p{1.2mm}p{1.2mm}|}
\hline
&\multicolumn{4}{c|}{T1}&\multicolumn{4}{c|}{T2}&\multicolumn{4}{c|}{T3}&\multicolumn{4}{c|}{T4}&\multicolumn{4}{c|}{T5}&\multicolumn{4}{c|}{T6}&\multicolumn{4}{c|}{T7}&\multicolumn{4}{c|}{T8}&\multicolumn{4}{c|}{T9}\\\hline
TP1&--&--&--&--&1&0&20&37&0&3&20&35&1&1&31&25&1&3&17&37&0&1&13&44&0&1&17&40&1&1&23&33&0&1&17&40\\\hline
TP2&1&0&20&37&--&--&--&--&0&1&10&47&0&1&16&41&0&0&15&43&0&2&16&40&0&0&9&49&0&0&12&46&0&0&10&48\\\hline
TP3&0&3&20&35&0&1&10&47&--&--&--&--&0&2&18&38&0&0&3&55&0&1&12&45&0&0&6&52&0&1&10&47&0&0&8&50\\\hline
TP4&1&1&31&25&0&1&16&41&0&2&18&38&--&--&--&--&0&2&12&44&0&2&20&36&0&0&13&45&0&0&14&44&0&0&14&44\\\hline
TP5&1&3&17&37&0&0&15&43&0&0&3&55&0&2&12&44&--&--&--&--&0&0&16&42&0&0&5&53&0&1&7&50&0&0&6&52\\\hline
TP6&0&1&13&44&0&2&16&40&0&1&12&45&0&2&20&36&0&0&16&42&--&--&--&--&0&0&10&48&0&0&19&39&0&0&9&49\\\hline
TP7&0&1&17&40&0&0&9&49&0&0&6&52&0&0&13&45&0&0&5&53&0&0&10&48&--&--&--&--&0&0&3&55&0&0&2&56\\\hline
TP8&1&1&23&33&0&0&12&46&0&1&10&47&0&0&14&44&0&1&7&50&0&0&19&39&0&0&3&55&--&--&--&--&0&0&2&56\\\hline
TP9&0&1&17&40&0&0&10&48&0&0&8&50&0&0&14&44&0&0&6&52&0&0&9&49&0&0&2&56&0&0&2&56&--&--&--&--\\\hline
\hline
Total&4&11&158&291&1&4&108&351&0&8&87&369&1&8&138&317&1&6&81&376&0&6&115&343&0&1&65&398&1&3&90&370&0&1&68&395\\\hline
\end{tabular}
\end{subtable}
\end{table*}

%% file: tables/time-sum.tex
\begin{table*}[t]
\centering
\scriptsize
\vspace{-0.25cm}
\hspace{-1cm}
\caption{Execution costs for the static TCP techniques. The table lists the average, min, max, and sum of costs across all subject programs for both test-class level and test-method level (i.e., cost at test-class level/cost at test-method level). Time is measured in second. \label{tab:time-sum}}
\vspace{-0.05cm}
\begin{tabular}{|c|cccc|cccc|}
\hline
\multirow{2}{*}{Techniques}&\multicolumn{4}{c|}{Pre-processing}&\multicolumn{4}{c|}{Test Prioritization}\\
&Avg.&Min&Max&Sum&Avg.&Min&Max&Sum\\
\hline
TP$_{cg-tot}$&244.14/244.14&1.21/1.21&13785.86/13785.86&14159.97/14159.97&0.20/0.41&0/0&3.10/10.58&11.37/23.78\\
TP$_{cg-add}$&244.14/244.14&1.21/1.21&113785.86/13785.86&14159.97/14159.97&0.18/0.64&0/0&2.87/19.98&10.59/37.02\\
TP$_{str}$&0.35/0.37&0.04/0.04&2.95/2.41&20.04/21.63&4.03/1,359.24&0.01/0.02&115.82/57,134.30&233.76/78,835.97\\
TP$_{topic-r}$&0.41/1.55&0.03/0.09&3.81/14.80&24.99/89.63&0.15/832.95&0/0.01&1.72/40,594.66&8.50/48,310.93\\
TP$_{topic-m}$&1.51/3.79&0.13/0.22&12.10/50.14&87.76/219.93&0.19/268.51&0/0.07&1.98/10,925.26&10.95/15,573.71\\\hline
\end{tabular}
\end{table*}

%% file: threats.tex
\vspace{-0.3cm}
\section{Threats to Validity}
\label{sec:threats}

\noindent
\textbf{Threats to Internal Validity:}
In our implementation, we used PIT to generate mutation faults to simulate real program faults. One potential threat is that the mutation faults may not reflect all ``natural" characteristics of real faults. However, mutation faults have been widely used in the domain of software engineering research and have, under proper circumstances, been demonstrated to be representative of the actual program faults~\cite{Just:FSE14}. Further threats related to mutation testing
include the potential bias introduced by equivalent and trivial mutants. In the context of our experimental settings, equivalent mutants will not be detected by test cases. As explained in Section \ref{sec:study}, we ignore all mutants that cannot be detected by test cases. Thus, we believe that this threat is sufficiently mitigated. To answer $RQ_1$-$RQ_3$, we randomly selected 500 faults (100 groups and five faults per group) for each subject system, which
may impact the evaluation of TCP performance. However, this follows the guidelines and methodology of previous
studies \cite{zhang2013bridging,Lu:ICSE16}, minimizing this threat. Additionally, we also introduce two research
questions, \textbf{$RQ_4$} and \textbf{$RQ_5$}, to investigate the impact of mutant quantities and type on TCP evaluation, allowing us to examine the validity of past evaluations of TCP effectiveness. There is also a potential threat due to trivial/subsumed mutants (e.g., those that are easily distinguished from the
original program) outlined in recent work \cite{Henard:ICSE16,Papadakis:ISSTA16}. The trivial or subsumed mutants may potentially impact the results (e.g., inflate the APFD values). However, we do not specifically control the trivial and subsumed mutants, following the body of previous work in the TCP area \cite{zhang2013bridging,Mei:TSE12,Lu:ICSE16}, since in practice real faults may also be trivial or subsume each other. In addition, we involve a large set of randomly selected mutants, further mitigating this threat to validity. We encourage future studies to further examine this potential threat.

To perform this study we reimplemented eight TCP techniques presented in prior work. It is possible that there may be some slight differences between the original authors' implementations and our own. However, we performed this task closely following the technical details of the prior techniques and set parameters following the guidelines in the original publications.  Additionally, the authors of this paper met for and open code review regarding the studied approaches. Furthermore, based on our general findings, we believe our implementations to be accurate.

\noindent
\textbf{Threats to External Validity:} The main external threat to our study is that we experimented on 58 software systems, which may impact the generalizability of the results. Involving more subject programs would make it easier to reason about how the studied TCP techniques would perform on software systems of different languages and purposes. However, we chose 58 systems with varying sizes (1.2KLoC - 83.0 KLoC) and different numbers of detectable mutants (132 - 46,429), which makes for a highly representative set of Java programs, more so than any past study. Additionally, some subjects were used as benchmarks in recent papers~\cite{Saha:ICSE15}. Thus, we believe our study parameters have sufficiently mitigated this threat to a point where useful and actionable conclusions can be drawn in the context of our research questions. In addition, we seeded mutants using operators provided in PIT. It is possible that having different types of operators or using different mutation analysis tool may impact the results of our study. However, PIT is one of the most popular mutation analysis tools and has been widely used in software testing research. Thus, we believe our design of the study has mitigated this threat. \blue{Finally, while it would be interesting to investigate the effectiveness of TCPs on detecting real regression faults, this is a difficult task.  A large set of real regression faults is notoriously hard to collect in practice. This is mainly due to the fact that during real-world software development, the developers usually run regression tests before committing new revisions to the code repositories, and will fix any regression faults before the commits, leaving few real regression faults recorded in the code repositories. On the other hand, mutants have been shown to be suitable for simulating real faults for software-testing experimentation~\cite{Just:FSE14, Andrews:ICSE05} (including test-prioritization experimentation~\cite{Do:06}). Furthermore, mutation testing is widely used in recent TCP research work \cite{Lu:ICSE16,Henard:ICSE16}. Thus, in this paper we evaluate TCP effectiveness in terms of detecting mutants, and leave the investigation of TCP performance on real regression faults as future work.}

Finally, we selected four static TCP techniques to experiment with in our empirical study. There are some other recent works proposing static TCP techniques~\cite{Arafeen:ICST13,Saha:ICSE15}, but we focus only on those which do not require additional inputs, such as code changes or requirements in this empirical study. Also, we only compared the static techniques with four state-of-art dynamic TCP techniques with statement-level coverage. We do not study the potential impact of different coverage granularities on dynamic TCPs. However, these four techniques are highly representative of dynamic techniques and have been widely used in TCP evaluation \cite{Lu:ICSE16, Rothermel:99, Do:04}, and statement-level coverage has been shown to be \textit{at least as effective} as other coverage types \cite{Lu:ICSE16}.

%% file: lessons.tex
\vspace{-0.3cm}
\section{Lessons Learned}
\vspace{-0.1cm}
In this section we comment on the lessons learned from this study and their potential impact on future research:

\noindent

\noindent
\textbf{Lesson 1:}
Our study illustrates that different test granularities impact the effectiveness of TCP techniques, and that the finer, method-level, granularity achieves better performance in terms of \apfd{} and \apfdc{}, detecting regression faults more quickly.  This finding should encourage researchers and practitioners to use method-level granularity, and perhaps explore even finer granularities for regression test-case prioritization. Additionally, researchers should evaluate their newly proposed approaches on different test granularities to better understand the effectiveness of new approaches. Moreover, APFDc values are relatively less consistent with APFD values at test-method level and vary more dramatically as compared to AFPD values. This suggests that researchers should evaluate novel TCP approaches in terms of different types of metrics to better investigate the approaches' effectiveness.

\noindent
\textbf{Lesson 2:} The performance of different TCPs varies across different subject programs. One technique may perform better on some subjects but perform worse on other subjects. For example, TP$_{topic}$ performs better than TP$_{cg-add}$ on $webbit$, but performs worse than TP$_{cg-add}$ on $wsc$. This finding suggests that the characteristics of each subject are important to finding suitable TCPs. Furthermore, we find that the selection of subject programs and the selection of implementation tools may carry a large impact regarding the results of the evaluation for TCPs (e.g., there can be large variance in the performance of different techniques depending on the subject, particularly for static approaches). This finding illustrates that the researchers need to evaluate their newly proposed techniques on a \textit{large} set of real subject programs to make their evaluation reliable. To facilitate this we provide links to download our subject programs and data at \cite{Qi:TSE17}. Additionally, a potential avenue for future research may be an adaptive TCP technique that is able to analyze certain characteristics of a subject program (e.g., complexity, test suite size, libraries used) and modify the prioritization technique to achieve peak performance.

\noindent
\textbf{Lesson 3:}
\blue{Our study demonstrates that while TCP techniques tend to perform better on larger programs, subject size does not significantly impact comparative measures of \apfd{} and \apfdc{} between TCP techniques. Thus, when the performance of TCP techniques are compared against each other on either large or small programs, similar results can be expected. This finding illustrates scalability of various TCP techniques. Also, our experimental results show that software evolution does not have clear impact on comparative TCP effectiveness.}

\noindent
\textbf{Lesson 4:}
Our study demonstrates that mutant quantity and type selected in the experimental settings for measuring the effectiveness of TCP techniques does not dramatically impact the results in terms of \apfd{} or \apfdc{} metrics. This finding provides practical guidelines for researchers, confirming the comparative validity of the mutant seeding process of prior TCP work, and also provides evidence that the fault quantity and type factors are less important to investigate in future work.

\noindent
\textbf{Lesson 5:} Our findings illustrate that the studied static and dynamic TCP techniques agree on only a small number of found faults for the top ranked test-methods and classes ranked by the techniques, and the most highly prioritized test cases by different TCP techniques share similar capabilities in detecting different types of mutation faults. This suggests several relevant avenues for future research. For instance, (i) it may be useful to investigate specific TCP techniques to detect important faults faster when considering the fault severity/importance \cite{Elbaum:01,Varun:ICJST10,Kavitha:10} instead of fault types (e.g., different mutant types) during testing; (ii) differing TCP techniques could be used to target specific types of faults or even faults in specific locations of a program; and (iii) static and dynamic information could \textit{potentially} be combined in order to achieve higher levels of effectiveness. Furthermore, the similarity study performed in this paper has not been a core part of many TCP evaluations, and we assert that such an analysis should be encouraged moving forward. While APFD and \apfdc{} provide a clear picture of the relative effectiveness of techniques, it cannot effectively illustrate \textit{the difference set} of detected faults between two techniques.  This is a critical piece of information when attempting to understand new techniques and how they relate to existing research.

%% file: conclude.tex
\vspace{-0.25cm}
\section{Conclusion}
\vspace{-0.05cm}
\label{sec:conclude}

In this work, we perform an extensive study empirically comparing the effectiveness, efficiency, and similarity of detected faults for static and dynamic TCP techniques on 58 real-world Java programs mined from GitHub. The experiments were conducted at both test-method and test-class levels to understand the impact of different test granularities on the effectiveness of TCP techniques. The results indicate that the studied static techniques tend to outperform the studied dynamic techniques at the test-class level in terms of both \apfd{} and \apfdc{} metrics, whereas dynamic techniques tend to outperform the static techniques at test-method level in terms of \apfd{}. APFDc values are generally consistent with APFD values at test-class level but relatively less consistent with APFDc at test-method level. In addition, APFDc values vary more dramatically across all subject programs as compared to APFD values. \blue{We also observed that subject size, software evolution, and mutant quantities and types within each faulty group/version do not significantly impact comparative measures of TCP effectiveness.} Additionally, we found that the faults uncovered by static and dynamic techniques for the highest prioritized test cases uncover mostly dissimilar faults, which suggests promising avenues for future work. Finally, we found evidence suggesting that different TCP techniques tend to perform differently across subject programs, which suggests that certain program characteristics may be important when considering which type of TCP technique to use.

%% file: acknowledgments.tex
\vspace{-0.25cm}
\section*{Acknowledgments}
\vspace{-0.05cm}
We would like thank the anonymous TSE and FSE reviewers for their insightful comments that significantly improved the quality of a previous version of this paper and provided useful suggestions for extending the study as presented in this journal paper submission. This work is supported in part by the NSF CCF-1218129, CCF-1566589, and NSF CNS-1510239 grants. Any opinions, findings, and conclusions expressed herein are the authors' and do not necessarily reflect those of the sponsors.